\documentclass[a4paper,english]{article}
\usepackage[T1]{fontenc}
\usepackage[latin9]{inputenc}
\usepackage{amsmath}
\usepackage{graphicx}
\usepackage{amssymb}
\usepackage{esint}
\usepackage[authoryear]{natbib}

\makeatletter

\newcommand{\lyxline}[1][1pt]{%
  \par\noindent%
  \rule[.5ex]{\linewidth}{#1}\par}

\usepackage{amsfonts}\@ifundefined{definecolor}
 {\usepackage{color}}{}
\usepackage{epsfig}\usepackage{fullpage}\usepackage{subfigure}\usepackage{fullpage}
\providecommand{\U}[1]{\protect\rule{.1in}{.1in}}
\newtheorem{theorem}{Theorem}\newcounter{hypA}\newenvironment{condition}{\refstepcounter{hypA}\begin{itemize}
\item[({\bf A\arabic{hypA}})]}{\end{itemize}}\newtheorem{proposition}{Proposition}\newtheorem{remark}{Remark}

\makeatother

\usepackage{babel}

\begin{document}

\title{Efficient Bayesian Inference for Switching State-Space Models using
Discrete Particle Markov Chain Monte Carlo Methods}

\author{Nick Whiteley, Christophe Andrieu\\
Department of Mathematics,\\
University of Bristol,\\
University Walk,\\
Bristol BS8 1TW, UK.\\
Email: \texttt{\{Nick.Whiteley,C.Andrieu\}@bris.ac.uk} \and Arnaud
Doucet\\
Department of Statistics,\\
University of British Columbia,\\
Vancouver V6T 1Z2, BC, Canada.\\
Email: \texttt{Arnaud@stat.ubc.ca}}
\maketitle
\begin{abstract}
Switching state-space models (SSSM) are a very popular class of time
series models that have found many applications in statistics, econometrics
and advanced signal processing. Bayesian inference for these models
typically relies on Markov chain Monte Carlo (MCMC) techniques. However,
even sophisticated MCMC methods dedicated to SSSM can prove quite
inefficient as they update potentially strongly correlated discrete-valued
latent variables one-at-a-time \citep{carterkohn1996,gerlach2000,giordani2008}.
Particle Markov chain Monte Carlo (PMCMC)\ methods are a recently
developed class of MCMC\ algorithms which use particle filters to
build efficient proposal distributions in high-dimensions \citep{andrieudoucetholenstein2008}.
The existing PMCMC methods of \citet{andrieudoucetholenstein2008}
are applicable to SSSM, but are restricted to employing standard particle
filtering techniques. Yet, in the context of discrete-valued latent
variables, specialised particle techniques have been developed which
can outperform by up to an order of magnitude standard methods \citep{fearnhead1998,fearnhead2003,Fearnhead2004}.
In this paper we develop a novel class of PMCMC\ methods relying
on these very efficient particle algorithms. We establish the theoretical
validy of this new generic methodology referred to as discrete PMCMC
and demonstrate it on a variety of examples including a multiple change-points
model for well-log data and a model for U.S./U.K. exchange rate data.
Discrete PMCMC\ algorithms are shown to outperform experimentally
state-of-the-art MCMC\ techniques for a fixed computational complexity.
Additionally they can be easily parallelized \citep{lee2009} which
allows further substantial gains.

\end{abstract}
\newpage{}

\section{Introduction\label{sec:intro}}

Linear Gaussian Switching State-Space Models (SSSM) are a class of
time series models in which the parameters of a linear Gaussian model
switch according to a discrete latent process. They are ubiquitous
in statistics \citep{cappe2005,fruwirthschnatter2006}, econometrics
\citep{kim1999,giordani2007} and advanced signal processing \citep{Barembauch08,costa2005}
as they allow us to describe in a compact and interpretable way regime
switching time series. SSSM have been successfully used to describe,
among others, multiple change-point models \citep{fearnhead2003,giordani2008},
nonparametric regression models with outliers \citep{carterkohn1996}
and Markov switching autoregressions \citep{billio1998,fruwirthschnatter2006,kim1999}.

Performing Bayesian inference for SSSM requires the use of Markov
chain Monte Carlo (MCMC) techniques. The design of efficient sampling
techniques for this class of models has been a subject of active research
for over fifteen years, dating back at least as far as \citet{carterkohn1994,shephard1994}.
A recent overview of MCMC in this context can be found in \citet{cappe2005,fruwirthschnatter2006}.
The main practical difficulty lies in simulating from the conditional
distribution of the trajectory of the discrete-valued latent process.
The cost of computing this distribution grows exponentially in the
length of the observation record and therefore obtaining an exact
sample from it is impractical for all but tiny data sets. A standard
strategy is instead to update the components of the discrete latent
process one-at-a-time \citep{carterkohn1996,gerlach2000,giordani2008}.
However, it is well-known that such an approach can significantly
slow down the convergence of MCMC\ algorithms. An alternative is
to sample approximately from the joint distribution of the latent
discrete trajectory using particle filters: non-iterative techniques
based on a combination of importance sampling and resampling techniques,
see \citet{doucet2001,Liu2001} for a review of the literature. Empirical
evidence suggests that particle filters are able to provide samples
whose distribution is close to the target distribution of interest
and this evidence is backed up by the rigourous quantitative bounds
established in \citet[chapter 8]{Delm04}. This motivates using particle
filters as proposal distributions within MCMC.

This idea is very natural, but its realization is far from trivial
as the distribution of a sample generated by a particle filter does
not admit a closed-form expression hence preventing us\ from directly
using the standard Metropolis-Hastings (MH) algorithm. In a recent
paper \citet{andrieudoucetholenstein2008} have shown that it is possible
to bypass this problem. The authors have proposed a whole class of
MCMC\ algorithms named Particle MCMC\ (PMCMC) relying on proposals
built using particle filters. These algorithms have been demonstrated
in the context of non-linear non-Gaussian state-space models and are
directly applicable to SSSM; see also \citet{flury2008} for applications
in financial econometrics. However, the standard particle methods
employed in \citet{andrieudoucetholenstein2008} do not fully exploit
the discrete nature of the latent process in SSSM. This was recognized
early by Paul Fearnhead who proposed an alternative generic algorithm,
which we refer to as the Discrete Particle Filter (DPF) \citep{fearnhead1998}.
The DPF bypasses the importance sampling step of standard particle
techniques and can be interpreted as using a clever random pruning
mechanism to select support points from the exponentially growing
sequence of discrete latent state spaces. The DPF methodology has
been demonstrated successfully in a variety of applications \citep{cappe2005,fearnhead1998,fearnhead2003,Fearnhead2004}.
It has been shown to significantly outperform alternative sophisticated
approaches such as the Rao-Blackwellized particle filters developed
in \citet{chenliu2000,doucet2000,doucet2001b} by up to an order of
magnitude for a fixed computational complexity.

The main contribution of this article is to propose a novel class
of PMCMC algorithms referred to as discrete PMCMC methods relying
on the DPF for this important class of statistical models. The practical
efficiency of the proposed methods relies on an original backward
sampling procedure. We show that on a variety of applications this
new generic methodology outperforms state-of-the-art MCMC\ algorithms
for a fixed computational complexity. Moreover, as in the case of
standard particle filters \citep{lee2009}, the DPF can be parallelized
easily. This suggests that even greater computational gains can be
achieved. 

The rest of the paper is organised as follows. In Section \ref{sec:discretevaluedlatent}
we present the general class of SSSM considered in this paper and
give an illustrative example. In Section \ref{sec:reviewMC}, we discuss
the intractability of exact inference in SSSM and present the DPF
algorithm \citep{fearnhead1998,fearnhead2003,Fearnhead2004}. Our
presentation is slightly non-standard and explicitly introduces the
random support sets generated by the algorithm. This allows us to
describe the DPF precisely and compactly in a probabilistic way which
proves useful to establish the validity of the proposed algorithms.
We also review standard MCMC\ techniques used in this context. In
Section \ref{sec:pmcmc} we introduce discrete PMCMC algorithms relying
on the DPF to perform inference in SSSM and present some theoretical
results. In Section \ref{sec:applications}, we review generic practical
issues and demonstrate the efficiency of the proposed methods in the
context of three examples. Finally in Section \ref{sec:discussion_extensions}
we discuss several extensions of this work.

\section{Switching state-space models\label{sec:discretevaluedlatent}}

\subsection{Model}

From herein, we use the standard convention whereby capital letters
are used for random variables while lower case letters are used for
their values. Hereafter for any generic process $\left\{ z_{n}\right\} $
we will denote $z_{i:j}:=\left(z_{i},z_{i+1},\ldots,z_{j}\right)$.
The identity matrix of size $p$ is denoted $I_{p}$ and the matrix
of zeros of size $p\times q$ by $0_{p\times q}$.

Consider the following SSSM, also known in the literature as a conditionally
linear Gaussian state-space model or a jump linear system. The latent
state process $\left\{ X_{n}\right\} _{n\geq1}$ is such that $X_{n}$
takes values in a \textbf{finite} set $\mathcal{X}$. It is characterized
by its initial distribution $X_{1}\sim\nu_{\theta}\left(\cdot\right)$
and transition probabilities for $n>1$\begin{equation}
X_{n}|(X_{1:n-1}=x_{1:n-1})\sim f_{\theta}\left(\cdot|x_{1:n-1}\right).\label{eq:evol}\end{equation}
 Conditional upon $\left\{ X_{n}\right\} _{n\geq1}$, we have a linear
Gaussian state-space model defined through $Z_{0}\sim\mathcal{N}(m_{0},\Sigma_{0})$
and for $n\geq1$ \begin{align}
Z_{n} & =A_{\theta}(X_{n})Z_{n-1}+B_{\theta}(X_{n})V_{n}+F_{\theta}(X_{n})u_{n},\text{ }\label{eq:evollineargauss}\\
Y_{n} & =C_{\theta}(X_{n})Z_{n}+D_{\theta}(X_{n})W_{n}+G_{\theta}(X_{n})u_{n},\label{eq:obslineargauss}\end{align}
where $\mathcal{N}(m,\Sigma)$ is the normal distribution of mean
$m$ and covariance $\Sigma$, $V_{n}\overset{\text{i.i.d.}}{\sim}\mathcal{N}(0_{v\times1},I_{v})$,
$W_{n}\overset{\text{i.i.d.}}{\sim}\mathcal{N}(0_{w\times1},I_{w})$,
$\left\{ A_{\theta}(x),B_{\theta}(x),C_{\theta}(x),D_{\theta}(x),F_{\theta}(x),G_{\theta}(x);x\in\mathcal{X}\right\} $
are matrices of appropriate dimension and $u_{n}$ is an exogeneous
input. Here $\theta\in\Theta$ is some static parameter which may
be multidimensional, for example $\Theta\subset\mathbb{R}^{d}$. For
purposes of precise specification of resampling algorithms in the
sequel and without loss of generality we label the elements of $\mathcal{X}$
with numbers, for example $\mathcal{X}=\left\{ 1,...,\mathcal{\left|X\right|}\right\} $
for some $\left|\mathcal{X}\right|\in\mathbb{N}$. We may then endow
each Cartesian product space $\mathcal{X}^{2},\mathcal{X}^{3},...$
with the corresponding lexicographical order relation. From henceforth,
whenever we refer to ordering of a set of points in $\mathcal{X}^{n}$
it is with respect to the latter relation.

We give here a simple example of a SSSM. Two more sophisticated examples
are discussed in Section \ref{sec:applications}.

\subsubsection{Example: Auto-regression with shifting level \label{subsec:Example:-Auto-regression-with}}

Let $\mathcal{X}=\left\{ 0,1\right\} $ and for $\left\{ X_{n}\right\} $
a Markov chain on $\mathcal{X}$ with transition matrix $P_{X}$,
consider the process defined by

\begin{eqnarray*}
Y_{n} & = & \mu_{n}+\phi(Y_{n-1}-\mu_{n-1})+\sigma V_{n,1}\\
\mu_{n} & = & \mu_{n-1}+\sigma X_{n}V_{n,2},\end{eqnarray*}
where for each $n\geq1$, $\mu_{n}$ and $Y_{n}$ are real-valued
and $\left\{ V_{n,1}\right\} $ and $\left\{ V_{n,2}\right\} $ are
i.i.d. $\mathcal{N}(0,1)$. The initial distribution on $\mu_{0}$
is $\mathcal{N}(m_{0},\sigma_{0}^{2})$ and is assumed known. This
is a natural generalization of a first order autoregressive model
to the case where the level $\mu_{n}$ is time-varying with shifts
driven by the latent process $\left\{ X_{n}\right\} $. This model
can be expressed in state-space form by setting\[
Z_{n}=\left[\begin{array}{c}
Y_{n}-\mu_{n}\\
\mu_{n}\end{array}\right],\quad A_{\theta}(x_{n})=\left[\begin{array}{cc}
\phi & 0\\
0 & 1\end{array}\right]\quad\forall x_{n},\]
\[
B_{\theta}(x_{n})=\sigma\left[\begin{array}{cc}
1 & 0\\
0 & x_{n}\end{array}\right],\quad C_{\theta}(x_{n})=\left[\begin{array}{cc}
1 & 1\end{array}\right],\quad D_{\theta}(x_{n})=F_{\theta}(x_{n})=G_{\theta}(x_{n})=0,\quad\forall x_{n}.\]
The unknown parameters of this model are $\theta=[\phi\;\sigma^{2}\; P_{X}]$
where $P_{X}$ is the transition matrix of $\{X_{n}\}$. In this model
and more generally in SSSMs, inferences about the latent processes
$\left\{ \mu_{n}\right\} $ and $\left\{ X_{n}\right\} $ from a particular
data set are likely to be highly sensitive to values of these parameters
if they are assumed known.

\subsection{Inference aims\label{subsec:inferenceaims}}

Our aim is to perform Bayesian inference in SSSMs, conditional upon
some observations $y_{1:T}$ and for some $T\geq1$, treating both
the latent trajectories $X_{1:T},Z_{0:T}$ and the parameter $\theta$
as unknowns. Where applicable, the values of the input sequence $u_{1:T}$
are assumed known, but for clarity we suppress them from our notation.
We ascribe a prior density $p\left(\theta\right)$ to $\theta$ so
Bayesian inference relies on the joint density \begin{equation}
p\left(\theta,x_{1:T},z_{0:T}|y_{1:T}\right)\propto p_{\theta}\left(x_{1:T},z_{0:T},y_{1:T}\right)p\left(\theta\right),\label{eq:jointposteriorstatespace}\end{equation}
where the definition of $p_{\theta}\left(x_{1:T},z_{0:T},y_{1:T}\right)$
follows from Eq. (\ref{eq:evol})-(\ref{eq:evollineargauss})-(\ref{eq:obslineargauss}).
This posterior can be factorized as follows \begin{equation}
p\left(\theta,x_{1:T},z_{0:T}|y_{1:T}\right)=p\left(\theta,x_{1:T}|y_{1:T}\right)p_{\theta}\left(z_{0:T}|y_{1:T},x_{1:T}\right)\label{eq:factorizationjoint}\end{equation}
 where \begin{equation}
p\left(\theta,x_{1:T}|y_{1:T}\right)=\dfrac{p_{\theta}\left(y_{1:T}|x_{1:T}\right)p(x_{1:T}|\theta)p(\theta)}{\int_{\Theta}\sum_{x_{1:T}'\in\mathcal{X}^{T}}p_{\theta}\left(y_{1:T}|x_{1:T}'\right)p(x_{1:T}'|\theta)p\left(\theta\right)\mathrm{d}\theta}.\label{eq:marginalposterior}\end{equation}
Conditional upon $X_{1:T}=x_{1:T},$ Eq. (\ref{eq:evollineargauss})-(\ref{eq:obslineargauss})
define a linear Gaussian state-space model so it is possible to compute
efficiently the statistics of the conditional multivariate Gaussian
density $p_{\theta}\left(\left.z_{0:T}\right\vert y_{1:T},x_{1:T}\right)$
in Eq. (\ref{eq:factorizationjoint}) and the conditional marginal
likelihood $p_{\theta}\left(\left.y_{1:T}\right\vert x_{1:T}\right)$
in Eq. (\ref{eq:marginalposterior}) using Kalman techniques. For
example $p_{\theta}\left(\left.y_{1:T}\right\vert x_{1:T}\right)$
can be computed using the product of predictive densities \begin{equation}
p_{\theta}\left(\left.y_{1:T}\right\vert x_{1:T}\right)={\displaystyle \prod\limits _{n=1}^{T}}g_{\theta}\left(\left.y_{n}\right\vert y_{1:n-1},x_{1:n}\right)\label{eq:likelihooddecomposepredictive}\end{equation}
 where $y_{1:0}:=\varnothing$. The statistics of these Gaussian predictive
densities can be computed using the Kalman filter which is recalled
in Appendix \ref{sec:kalman} for sake of convenience. For simplicity
of presentation throughout the following we assume that for each $1\leq n\leq T$
and $\theta\in\Theta$ the support of $p_{\theta}\left(x_{1:n}\left|y_{1:n}\right.\right)$
is $\mathcal{X}^{n}$. This assumption is satisfied in the vast majority
of cases considered in practice and in all the examples we consider.
The techniques discussed below can be transferred to cases where this
assumption is not met with only cosmetic changes.

\section{Inference techniques for switching state-space models\label{sec:reviewMC}}

\subsection{Exact Inference and Intractability}

The main difficulty faced in the exact computation of $p(\theta,x_{1:T}|y_{1:T})$,
is the need to perform the summation in the denominator of Eq. \eqref{eq:marginalposterior}
over up to $\left\vert \mathcal{X}\right\vert ^{T}$ values of $x_{1:T}$,
where $\left\vert \mathcal{X}\right\vert $ is the cardinality of
$\mathcal{X}$. For even modest values of $T$, this sum is too expensive
to compute exactly. In the applications we consider, $T$ is of the
order of thousands, so exact computation is practically impossible.

Even if $\theta$ is treated as fixed, inference is intractable. In
this case, we wish to compute $p_{\theta}(x_{1:T}|y_{1:T})$, whose
normalization involves the same problematic summation. One approach
is to obtain $p_{\theta}(x_{1:T}|y_{1:T})$ by sequential computation
of $p_{\theta}(x_{1}|y_{1}),p_{\theta}(x_{1:2}|y_{1:2}),...$ via
the recursive relationship \[
p_{\theta}(x_{1:n}|y_{1:n})=\dfrac{g_{\theta}(y_{n}|y_{1:n-1},x_{1:n})f_{\theta}(x_{n}|x_{1:n-1})p_{\theta}(x_{1:n-1}|y_{1:n-1})}{\sum_{x_{1:n\in\mathcal{X}^{n}}}g_{\theta}(y_{n}|y_{1:n-1},x_{1:n})f_{\theta}(x_{n}|x_{1:n-1})p_{\theta}(x_{1:n-1}|y_{1:n-1})},\]
 (with $g_{\theta}(y_{n}|y_{1:n-1},x_{1:n})$ the predictive as defined
in the previous section) but the computation involved increases exponentially
in $n$. For purposes of exposition in the sequel, we remark that,
as for each $n$ the support of $p_{\theta}(x_{1:n}|y_{1:n})$ is
$\mathcal{X}^{n}$ then the sequence of such supports satisfies the
trivial recursion \[
\mathcal{X}^{n}=\mathcal{X}\times\mathcal{X}^{n-1},\]
and is evidently growing in cardinality with $n$. Hence, in both
the cases of computing $p(\theta,x_{1:T}|y_{1:T})$ and $p_{\theta}(x_{1:T}|y_{1:T})$
it is necessary to rely on approximations and we focus here on Monte
Carlo methods.

\subsection{Monte Carlo Methods}

We next review two classes of Monte Carlo techniques to perform inference
in SSSM. The first method we discuss is the DPF algorithm of \citet{fearnhead1998}.
For a fixed parameter value $\theta$, this algorithm allows us to
compute an approximation of the posterior distribution $p_{\theta}\left(x_{1:T}|y_{1:T}\right)$
and an approximation of the marginal likelihood $p_{\theta}\left(y_{1:T}\right)$.
We present this algorithm in a slightly non-standard way which allows
us to describe it probabilistically in a concise and precise manner.
This will prove useful for the development of the discrete PMCMC algorithms
in Section \ref{sec:pmcmc}. We also review MCMC methods which have
been developed to approximate $p\left(\theta,x_{1:T},z_{0:T}|y_{1:T}\right)$
and discuss their advantages and limitations.

\subsection{The discrete particle filter\label{sec:descriptionSMC}}

The DPF algorithm proposed in \citet{fearnhead1998,fearnhead2003}
is a non-iterative procedure approximating the posterior distribution
$p_{\theta}\left(x_{1:T}|y_{1:T}\right)$ and the marginal likelihood
$p_{\theta}\left(y_{1:T}\right)$. Practically, the DPF approximation
of the posterior distributions $\left\{ p_{\theta}\left(x_{1:n}|y_{1:n}\right);n\geq1\right\} $
is made sequentially in time using a collection of $N\left\vert \mathcal{X}\right\vert $
weighted trajectories or {}``particles\textquotedblright{}\ $\left\{ X_{1:n}^{\left(i\right)};i=1,...,N\left\vert \mathcal{X}\right\vert \right\} $,\[
\widehat{p}_{\theta}^{N}\left(x_{1:n}|y_{1:n}\right)=\sum_{i=1}^{N\left\vert \mathcal{X}\right\vert }W_{n}^{\theta}\left(X_{1:n}^{\left(i\right)}\right)\delta_{X_{1:n}^{\left(i\right)}}\left(x_{1:n}\right),\text{ }W_{n}^{\theta}\left(X_{1:n}^{\left(i\right)}\right)\geq0,\text{ }\sum_{i=1}^{N\left\vert \mathcal{X}\right\vert }W_{n}^{\theta}\left(X_{1:n}^{\left(i\right)}\right)=1.\]
The parameter $N$ controls the precision of the algorithm. The larger
it is, the more accurate (on average) the approximation of the target
distribution. It has been demonstrated experimentally in \citet{cappe2005,fearnhead2003,Fearnhead2004}
that the DPF algorithm outperforms significantly, sometimes by one
order of magnitude, the Rao-Blackwellized particle filters proposed
in \citet{chenliu2000,doucet2000,doucet2001b} and that it is able
to provide very good approximations of $p_{\theta}\left(x_{1:T}|y_{1:T}\right)$
in realistic scenarios even with a moderate number of particles. The
action of the DPF can be summarised as follows.

Assume that we have, at time step $n$ obtained $\widehat{p}_{\theta}^{N}\left(x_{1:n}|y_{1:n}\right)$
consisting of $N\left\vert \mathcal{X}\right\vert $ distinct particles
with weights that sum to 1. A resampling step is then applied, exactly
$N$ of the $N|\mathcal{X}|$ trajectories survive and their weights
are adjusted accordingly. The resampling mechanism is chosen in such
a way as to be optimal in some sense. Throughout the remainder of
the paper we treat the case of minimising the sum of variances of
the importance weights as in \citet{fearnhead2003} but exactly the
same method applies to other schemes discussed in \citet{Barembauch08}.
Features of this resampling scheme which distinguish it from standard
methods, such as multinomial resampling, are that it results in no
duplicated particles and gives post-resampling weights which are non-uniform.

Whereas standard particle methods rely on a stochastic proposal mechanism
to explore the space, the DPF performs all its exploration deterministically.
This is possible because of the finite cardinality of the latent discrete
space. Consider one of $N$ particles which survived the resampling
operation, each of which is a point in $\mathcal{X}^{n}$. Call the
point in question $x_{1:n}$ and denote by \textsf{$m_{n|n}^{z,\theta}(x_{1:n})$}
and\textsf{ $\Sigma_{n|n}^{z,\theta}(x_{1:n})$} respectively the
mean and covariance of the Gaussian density $p_{\theta}(z_{n}|y_{1:n},x_{1:n})$.
From this point $|\mathcal{X}|$ new particles $\{(x_{1:n},x);x\in\mathcal{X}\}$
are formed, and for each one of them, \textsf{$m_{n+1|n+1}^{z,\theta}(x_{1:n},x)$},
\textsf{$\Sigma_{n+1|n+1}^{z,\theta}(x_{1:n},x)$} and the associated
unnormalized weight are calculated using the Kalman filtering recursions
(included for reference in Appendix \ref{sec:kalman}). This procedure
is repeated for the remaining $N-1$ particles, resulting in $N|\mathcal{X}|$
weighted trajectories. The weights are then normalized to yield a
probability distribution constituting $\widehat{p}_{\theta}^{N}\left(x_{1:n+1}|y_{1:n+1}\right)$. 

This outline of the DPF operations highlights the function of the
resampling step: in the case of the DPF it acts to prune the exponentially
growing (in $n$) tree of possible paths $\left\{ x_{1:n}\in\mathcal{X}^{n};n=1,2,...\right\} $.
It is convenient to specify the DPF in a slightly non-standard way
which highlights that the only randomness in this algorithm arises
from the resampling step. To this end, we introduce random support
sets $\mathbf{S}_{1},\mathbf{S}_{2},...,\mathbf{S}_{T}$ with each
$\mathbf{S}_{n}$ taking a value $\mathbf{s}_{n}$ which is a subset
of $\mathcal{X}^{n}$. It is stressed that, in the following interpretation,
the $x_{1:n}$'s are not random variables, and are just points in
the state space (and Cartesian products thereof) used for indexing.
With this notation, we write the DPF approximation for $n>1$ as \begin{align}
\widehat{p}_{\theta}^{N}\left(x_{1:n}|y_{1:n}\right) & =\sum_{x_{1:n}^{\prime}\in\mathbf{S}_{n}}W_{n}^{\theta}\left(x_{1:n}^{\prime}\right)\delta_{x_{1:n}^{\prime}}\left(x_{1:n}\right).\label{eq:alternativerepresentationDPF}\end{align}
Under the probability law of the DPF algorithm, which we discuss in
more detail later, for each $n\geq2$, $|\mathbf{S}_{n}|=N|\mathcal{X}|\wedge\left|\mathcal{X}^{n}\right|$,
with probability $1$. We thus see in Eq. \eqref{eq:alternativerepresentationDPF}
the effect of the parameter $N$: it specifies the number of support
points of the approximation $\widehat{p}_{\theta}^{N}\left(x_{1:n}|y_{1:n}\right)$.
We next provide pseudo code for the DPF algorithm and then go on to
discuss several issues related to its practical use and its theoretical
representation.

\pagebreak{}

\lyxline{\normalsize}

\noindent \begin{center}
\textbf{DPF algorithm} 
\par\end{center}

\hspace{-0.5cm}\underline{\textsf{At time }$n=1$}

$\hspace{-0.5cm}\bullet$ \textsf{Set } $\mathbf{S}_{1}=\mathcal{X}$
\textsf{ and for each }$x_{1}\in\mathcal{X}$,\textsf{ compute $m_{1|1}^{z,\theta}(x_{1})$
, $\Sigma_{1|1}^{z,\theta}(x_{1})$ and $g_{\theta}(y_{1}|x_{1})$
using the Kalman filter. }

$\hspace{-0.5cm}\bullet$\textsf{ Compute and normalise the weights.
For each $x_{1}\in\mathcal{X}$, }\begin{equation}
\overline{w}_{1}^{\theta}\left(x_{1}\right)=\nu_{\theta}\left(x_{1}\right)g_{\theta}\left(\left.y_{1}\right\vert x_{1}\right),\text{ }W_{1}^{\theta}\left(x_{1}\right)=\frac{\overline{w}_{1}^{\theta}\left(x_{1}\right)}{{\textstyle \sum\nolimits _{x_{1}^{\prime}\in\mathcal{X}}}\overline{w}_{1}^{\theta}\left(x_{1}^{\prime}\right)}.\label{eq:unnormalizedweightfilteringinit}\end{equation}

\vspace{-0.25cm}
\hspace{-0.5cm}\underline{\textsf{At times
}$n=2,...,T$}

$\hspace{-0.5cm}\bullet$ \textsf{If $\left|\mathbf{S}_{n-1}\right|\leq N$
set $C_{n-1}=\infty$ otherwise set }$C_{n-1}$\textsf{ to the unique
solution of} \[
\sum_{x_{1:n-1}\in\mathbf{S}_{n-1}}1\wedge C_{n-1}W_{n-1}^{\theta}\left(x_{1:n-1}\right)=N.\]

$\hspace{-0.5cm}\bullet$ \textsf{Maintain the }$L_{n-1}$\textsf{
trajectories in } $\mathbf{S}_{n-1}$ \textsf{ which have weights
strictly superior to }$1/C_{n-1}$\textsf{, then apply the stratified
resampling mechanism to the other }$N\left|\mathcal{X}\right|-L_{n-1}$\textsf{
trajectories to yield }$N-L_{n-1}$\textsf{ survivors. Set} $\mathbf{S}_{n-1}'$
\textsf{ to the set of surviving and maintained trajectories.}

$\hspace{-0.5cm}\bullet$ \textsf{Set }$\mathbf{S}_{n}=\mathbf{S}_{n-1}'\times\mathcal{X}$.

$\hspace{-0.5cm}\bullet$ \textsf{For each $x_{1:n}\in\mathbf{S}_{n}$,
compute $m_{n|n}^{z,\theta}(x_{1:n})$ , $\Sigma_{n|n}^{z,\theta}(x_{1:n})$
and $g_{\theta}(y_{n}|y_{1:n-1},x_{1:n})$ using the Kalman filter.}

$\hspace{-0.5cm}\bullet$ \textsf{Compute and normalise the weights.
For each $x_{1:n}\in\mathbf{S}_{n},$ }\begin{align}
\overline{w}_{n}^{\theta}\left(x_{1:n}\right) & =f_{\theta}(x_{n}|x_{1:n-1})g_{\theta}(y_{n}|y_{1:n-1},x_{1:n})\dfrac{W_{n-1}^{\theta}\left(x_{1:n-1}\right)}{1\wedge C_{n-1}W_{n-1}^{\theta}\left(x_{1:n-1}\right)},\label{eq:updateweight1}\\
W_{n}^{\theta}\left(x_{1:n}\right) & =\frac{\overline{w}_{n}^{\theta}\left(x_{1:n}\right)}{{\textstyle \sum\nolimits _{x_{1:n}^{\prime}\in\mathbf{S}_{n}}}\overline{w}_{n}^{\theta}\left(x_{1:n}^{\prime}\right)}.\label{eq:updateweight2}\end{align}

\lyxline{\normalsize}

\subsubsection{Exact computation at the early iterations }

For small $n$ it is practically possible to compute $p_{\theta}(x_{1:n}|y_{1:n})$
exactly. It is only once $n$ is large enough that $\left|\mathcal{X}^{n}\right|>N$
that we need to employ the resampling mechanism to prune the set of
trajectories. This action is represented conceptually in the DPF algorithm
above by the artifice of setting $C_{n}=\infty$ if $n$ is such that
$\left|\mathbf{S}_{n-1}\right|\leq N$. When this condition is satisfied,
the resampling step is not called into action. Of course in the practically
unrealistic case that $\left|\mathcal{X}^{T}\right|\leq N$ the DPF,
unlike standard SMC algorithms, thus reduces to exact recursive computation
of $\left\{ p_{\theta}(x_{1:n}|y_{1:n});n=1,...,T\right\} $.

\subsubsection{Computing $C_{n}$ and stratified resampling}

The threshold $C_{n}$ is a deterministic function of the weights
$\left\{ W_{n}^{\theta}\left(x_{1:n}\right)\right\} _{x_{1:n}\in\mathbf{S}_{n}}$.
A method for solving $\sum_{x_{1:n}\in\mathbf{S}_{n}}1\wedge C_{n}W_{n}^{\theta}\left(x_{1:n}\right)=N$
is given in \citet{fearnhead2003}. The stratified resampling mechanism,
which is employed once $C_{n}$ has been computed, proceeds as follows
at time $n$; this was originally proposed in \citet{carpenter1999,kitagawa1996},
although not in the context of the DPF.

\lyxline{\normalsize}

\noindent \begin{center}
\textbf{Stratified resampling} 
\par\end{center}

$\hspace{-0.5cm}\bullet$ \textsf{Normalise the weights }$\overline{w}_{n-1}^{\theta}\left(x_{1:n-1}\right)$\textsf{
of the }$N\left|\mathcal{X}\right|-L_{n-1}$\textsf{ particles and
label them according to the order of the corresponding $x_{1:n-1}$
to obtain }$\widehat{W}_{n-1}^{\theta}\left(x_{1:n-1}^{\left(i\right)}\right)$\textsf{;}
$i=1,...,N\left|\mathcal{X}\right|-L_{n-1}.$

$\hspace{-0.5cm}\bullet$\textsf{ Construct the corresponding cumulative
distribution function: for} $i=1,...,N\left|\mathcal{X}\right|-L_{n-1}$,

\[
Q_{n-1}^{\theta}(i):=\sum_{j\leq i}\widehat{W}_{n-1}^{\theta}\left(x_{1:n-1}^{\left(j\right)}\right),\qquad Q_{n-1}^{\theta}(0):=0.\]

$\hspace{-0.5cm}\bullet$ \textsf{Sample }$U_{1}$ \textsf{uniformly
on }$\left[0,1/(N-L_{n-1})\right]$\textsf{ and set }$U_{j}=U_{1}+\frac{j-1}{N-L_{n-1}}$\textsf{
for }$j=2,...,N-L_{n-1}.$

$\hspace{-0.5cm}\bullet$ \textsf{For} $i=1,...,N\left|\mathcal{X}\right|-L_{n-1}$,
\textsf{if there exists} $j\in\left\{ 1,...,N-L_{n-1}\right\} $ \textsf{such
that} $Q_{n-1}^{\theta}(i-1)<U_{j}\leq Q_{n-1}^{\theta}(i)$, \textsf{then}
$x_{1:n-1}^{(i)}$ \textsf{survives}.

\lyxline{\normalsize}

\subsubsection{Computational Requirements }

\noindent Assuming that the cost of evaluating $f_{\theta}(x_{n}|x_{1:n-1})$
is $\mathcal{O}(1)$ for all $n$, the computational complexity of
the DPF is $\mathcal{O}(|\mathcal{X}|N)$ at each time step due to
the propagation of $N|\mathcal{X}|$ Kalman filtering operations and
the generation of a single uniform random variable. The parallelisation
techniques described in \citet{lee2009} could readily be exploited
when performing the Kalman computations.

\subsubsection{Estimating \textmd{$p_{\theta}\left(y_{1:T}\right)$}}

Of particular interest in the sequel is the fact that the DPF provides
us with an estimate of the marginal likelihood $p_{\theta}\left(y_{1:T}\right)$
given by \begin{equation}
\widehat{p}_{\theta}\left(y_{1:T}\right):=\widehat{p}_{\theta}\left(y_{1}\right){\displaystyle \prod\limits _{n=2}^{T}}\widehat{p}_{\theta}\left(y_{n}|y_{1:n-1}\right)\label{eq:marginallikelihoodSMC}\end{equation}
 where \begin{equation}
\widehat{p}_{\theta}\left(y_{1}\right)=\sum_{x_{1}\in\mathcal{X}}\overline{w}_{1}^{\theta}\left(x_{1}\right),\;\;\;\;\widehat{p}_{\theta}\left(y_{n}|y_{1:n-1}\right)=\sum_{x_{1:n}\in\mathbf{S}_{n}}\overline{w}_{n}^{\theta}\left(x_{1:n}\right),\;\; n>1.\label{eq:marginallikelihoodincrementsSMC}\end{equation}
Inevitably, for fixed $N$, the quality of the particle approximation
to the distribution $p_{\theta}\left(x_{1:T}|y_{1:T}\right)$ decreases
as $T$ increases. For fixed $T$, once $N$ is larger than $\left|\mathcal{X}^{T}\right|$,
the DPF computes $p_{\theta}\left(y_{1:T}\right)$ exactly.

Before introducing the details of the new PMCMC algorithms, we review
some existing MCMC algorithms for performing inference in SSSM.

\subsection{Standard Markov chain Monte Carlo methods\label{sec:descriptionMCMC}}

Designing efficient MCMC algorithms to sample from $p\left(\theta,x_{1:T},z_{0:T}|y_{1:T}\right)$
is a difficult task. Most existing MCMC methods approach this problem
using some form of Gibbs sampler and can be summarized as cycling
in some manner through the sequence of distributions $p\left(\theta|y_{1:T},x_{1:T},z_{0:T}\right)$,
$p_{\theta}\left(z_{0:T}|y_{1:T},x_{1:T}\right)$ and $p_{\theta}\left(x_{1:T}|y_{1:T},z_{0:T}\right)$
or $p_{\theta}\left(x_{1:T}|y_{1:T}\right)$.

Sampling efficiently from $p\left(\theta|y_{1:T},x_{1:T},z_{0:T}\right)$
is often feasible due to the small or moderate size of $\theta$ and
the fact that for many models and parameters of interest, conjugate
priors are available. When conjugate priors are not used, Metropolis-within-Gibbs
steps may be applied.

A variety of efficient algorithms have been developed to sample from
$p_{\theta}\left(z_{0:T}|y_{1:T},x_{1:T}\right)$. These methods rely
on the conditionally linear Gaussian structure of the model and involve
some form of forward filtering backward sampling recursion \citep{carterkohn1994,fruwirthschnatter1994}.
Variants of these schemes which approach the task by explicitly sampling
the state disturbances may be more efficient and/or numerically stable
for some classes of models \citep{dejong1995,durbin2002}. In all
the numerical examples we consider, sampling from $p_{\theta}\left(z_{0:T}|y_{1:T},x_{1:T}\right)$
was performed using the simulation smoother of \citet{durbin2002}.

Sampling from $p_{\theta}\left(x_{1:T}|y_{1:T},z_{0:T}\right)$ can
also be performed efficiently using a forward filtering backward sampling
recursion \citep{carterkohn1994,chib1996} when $\left\{ X_{n}\right\} $
is a Markov chain. The resulting Gibbs sampler is elegant but it can
mix very slowly as $X_{1:T}$ and $Z_{0:T}$ are usually strongly
correlated. To bypass this problem, \citet{carterkohn1996,gerlach2000}
proposed to integrate out $Z_{0:T}$ using the Kalman filter as discussed
in Subsection \ref{subsec:inferenceaims}. However, as mentioned in
the introduction, exact sampling from $p_{\theta}\left(x_{1:T}|y_{1:T}\right)$
is typically infeasible as the cost of computing this distribution
is exponential in $T$. Therefore, in the algorithms of \citet{carterkohn1996,gerlach2000},
the discrete variables $X_{1:T}$ are updated one-at-a-time according
to their full conditional distributions $p_{\theta}\left(x_{n}|y_{1:T},x_{1:n-1},x_{n+1:T}\right)$.
It was shown in \citet{carterkohn1996,gerlach2000} that this strategy
can improve performance drastically compared to algorithms where $X_{1:T}$
is updated conditional upon $Z_{0:T}$. From hereon we refer to the
Gibbs sampler of \citet{gerlach2000} as the ``standard Gibbs'' algorithm.

At this stage, we comment a little further on the method of \citet{gerlach2000}
as it is relevant to the new algorithms described in the later sections.
The Gibbs sampler of \citet{gerlach2000} achieves a sweep of samples
from $p_{\theta}\left(x_{1}|y_{1:T},x_{2:T}\right)$, $p_{\theta}\left(x_{2}|y_{1:T},x_{1},x_{3:T}\right)$,
etc. by a {}``backward--forward'' procedure exploiting the identities
\begin{equation}
p_{\theta}\left(x_{n}|y_{1:T},x_{1:n-1},x_{n+1:T}\right)\propto p_{\theta}(y_{n}|y_{1:n-1},x_{1:n})p_{\theta}(x_{n}|x_{1:n-1},x_{n+1:T})p_{\theta}(y_{n+1:T}|y_{1:n},x_{1:T}),\label{eq:gerlach_full_conditional}\end{equation}
 and \begin{equation}
p_{\theta}(y_{n+1:T}|y_{1:n},x_{1:T})=\int p_{\theta}(y_{n+1:T}|z_{n},x_{n+1:T})p_{\theta}(z_{n}|x_{1:n},y_{1:n})dz_{n}.\label{eq:gerlach_identity}\end{equation}
In \citet{gerlach2000}, it was shown that the coefficients of $z_{n}$
in $p_{\theta}(y_{n+1:T}|z_{n},x_{n+1:T})$ which are needed to evaluate
(\ref{eq:gerlach_identity}) can be computed recursively for $n=T,T-1,...,1$
(the backward step). Then, for each $n=1,2,...,T$, $p_{\theta}(y_{n}|y_{1:n-1},x_{1:n})$
and $p_{\theta}(z_{n}|x_{1:n},y_{1:n})$ are obtained through standard
Kalman filtering recursions, (\ref{eq:gerlach_identity}) is computed
for each $x_{n}\in\mathcal{X}$ and a draw is made from (\ref{eq:gerlach_full_conditional})
(the forward step). In the resulting algorithm, if the computational
cost of evaluating $p_{\theta}(x_{n}|x_{1:n-1},x_{n+1:T})$ is $\mathcal{O}(1)$,
the cost of one sampling sweep through $p_{\theta}\left(x_{1}|y_{1:T},x_{2:T}\right)$,
$p_{\theta}\left(x_{2}|y_{1:T},x_{1},x_{3:T}\right)$, etc. grows
linearly is $\mathcal{O}(T)$.

More recently, adaptive MCMC methods have been suggested to make one-at-a-time
updates \citep{giordani2008}. However, these algorithms are still
susceptible to slow mixing if the components of $X_{1:T}$ are strongly
correlated. Moreover even if we were able to sample efficiently using
one-at-a-time updates, this algorithm might still converge slowly
if $X_{1:T}$ and $\theta$ are strongly correlated; e.g. if $\left\{ X_{n}\right\} $
is a Markov chain and $\theta$ includes the transition matrix of
this chain. \citet{Ham2011} have suggested an approximate, deterministic
algorithm for forward filtering-backward smoothing in switching state
space models, and the use of this method for making independent proposals
as part of a Metropolis-Hastings scheme. By contrast, and as we shall
see in the following section, the stochastic nature of the DPF algorithm
allows the construction not only of exact Metropolis-Hastings-type
algorithms, but also exact Particle Gibbs samplers. It is not clear
how to achieve the latter using the deterministic forward-backward
algorithm of \citet{Ham2011}.

\section{Discrete particle Markov chain Monte Carlo methods for switching
state-space models\label{sec:pmcmc}}

A natural idea arising from the previous section is to use the output
$\widehat{p}_{\theta}\left(x_{1:T}|y_{1:T}\right)$ of the DPF algorithm
as part of a proposal distribution for a MCMC algorithm targeting
$p_{\theta}\left(x_{1:T}|y_{1:T}\right)$ or $p\left(\theta,x_{1:T}|y_{1:T}\right)$.
This could allow us, in principle, to design automatically an efficient
high-dimensional proposal for MCMC. However a direct application of
this idea would require us to be able to both sample from and evaluate
pointwise the \emph{unconditional} distribution of a particle sampled
from $\widehat{p}_{\theta}\left(x_{1:T}|y_{1:T}\right)$. This distribution
is given by \[
q_{\theta}\left(x_{1:T}|y_{1:T}\right)=\mathbb{E}\left[\widehat{p}_{\theta}\left(x_{1:T}|y_{1:T}\right)\right],\]
where the expectation is with respect to the probability law of the
DPF algorithm: the stochasticity which produces the random probability
measure $\widehat{p}_{\theta}\left(x_{1:T}|y_{1:T}\right)$ in Eq.
(\ref{eq:alternativerepresentationDPF}). While sampling from $q_{\theta}\left(x_{1:T}|y_{1:T}\right)$
is straightforward as it only requires running the DPF\ algorithm
to obtain $\widehat{p}_{\theta}\left(x_{1:T}|y_{1:T}\right)$ then
sampling from this random measure, the analytical expression for this
distribution is clearly not available.

The novel MCMC updates presented in this section, under the umbrella
term discrete PMCMC, circumvent this problem by considering target
distributions on an extended space, over all the random variables
of the DPF algorithm. Details of their theoretical validity are given
in Subsection \ref{sec:proofs} but are not required for implementation
of the algorithms. The key feature of these discrete PMCMC\ algorithms
is that they are {}``exact approximations\textquotedblright{} to
standard MCMC updates targeting $p\left(\theta,x_{1:T}|y_{1:T}\right)$.
More precisely, on the one hand these algorithms can be thought of
as approximations to possibly {}``idealized\textquotedblright{} standard
MH updates parametrized by the number $N$ of particles used to construct
the DPF approximation. On the other hand, under mild assumptions,
discrete PMCMC algorithms are guaranteed to generate asymptotically
(in the number of MCMC\ iterations used) samples from $p\left(\theta,x_{1:T}|y_{1:T}\right)$,
\emph{for any fixed number }$N\geq2$\emph{ of particles}, in other
words, for virtually any degree of approximation.

In Subsection \ref{sec:particlemmh}, we describe the \emph{Particle
MMH} (Marginal Metropolis-Hastings) algorithm which can be thought
of as an exact approximation of an idealised {}``Marginal MH\textquotedblright{}
(MMH) targeting directly the marginal distribution $p\left(\theta|y_{1:T}\right)$
of $p\left(\theta,x_{1:T}|y_{1:T}\right)$. This algorithm admits
a form similar to the PMMH discussed in \citet{andrieudoucetholenstein2008}
but its validity relies on different arguments. In Subsection \ref{sec:particlegibbs_for_SSM}
we present a particle approximation of a Gibbs sampler targeting $p\left(\theta,x_{1:T}|y_{1:T}\right)$,
called the \emph{Particle Gibbs} (PG) algorithm. It is a particle
approximation of the {}``ideal\textquotedblright{} block Gibbs sampler
which samples from $p\left(\theta,x_{1:T}|y_{1:T}\right)$ by sampling
iteratively from the full conditionals $p_{\theta}\left(x_{1:T}|y_{1:T}\right)$
and $p\left(\theta|y_{1:T},x_{1:T}\right)$. This algorithm is significantly
different from the PG sampler presented in \citet{andrieudoucetholenstein2008}
and incorporates a novel backward sampling mechanism. Convergence
results for these algorithms are established in Subsection \ref{sec:proofs}.

\subsection{Particle marginal Metropolis-Hastings sampler\label{sec:particlemmh}}

Let us consider the following ideal {}``marginal\textquotedblright{}
MH (MMH)\ algorithm to sample from $p\left(\theta,x_{1:T}|y_{1:T}\right)$
where $\theta$ and $x_{1:T}$ are updated simultaneously using the
proposal given by \[
q\left(\left.\left(\theta^{\ast},x_{1:T}^{\ast}\right)\right\vert \left(\theta,x_{1:T}\right)\right)=q\left(\left.\theta^{\ast}\right\vert \theta\right)p_{\theta^{\ast}}\left(x_{1:T}^{\ast}|y_{1:T}\right)\ .\]
 In this scenario the proposed $X{}_{1:T}^{\ast}$ is perfectly {}``adapted\textquotedblright{}\ to
the proposed $\theta^{\ast}$ and the resulting MH\ acceptance ratio
is given by\begin{equation}
\frac{p\left(\theta^{\ast},x_{1:T}^{\ast}|y_{1:T}\right)}{p\left(\theta,x_{1:T}|y_{1:T}\right)}\frac{q\left(\left.\left(\theta,x_{1:T}\right)\right\vert \left(\theta^{\ast},x_{1:T}^{\ast}\right)\right)}{q\left(\left.\left(\theta^{\ast},x_{1:T}^{\ast}\right)\right\vert \left(\theta,x_{1:T}\right)\right)}=\frac{p_{\theta^{\ast}}\left(y_{1:T}\right)\text{ }p\left(\theta^{\ast}\right)}{p_{\theta}\left(y_{1:T}\right)\text{ }p\left(\theta\right)}\frac{q\left(\theta|\theta^{\ast}\right)}{q\left(\theta^{\ast}|\theta\right)}\ .\label{eq:optimalacceptanceprobammh}\end{equation}
 This algorithm is equivalent to a MH update working directly on the
marginal density $p\left(\theta|y_{1:T}\right)$, justifying the MMH
terminology. This algorithm is appealing but typically cannot be implemented
as the marginal likelihood terms $p_{\theta}\left(y_{1:T}\right)$
and $p_{\theta^{\ast}}\left(y_{1:T}\right)$ cannot be computed exactly
and it is impossible to sample exactly from $p_{\theta^{\ast}}\left(x_{1:T}|y_{1:T}\right)$.
We propose the following particle approximation of the MMH\ algorithm
where, whenever a sample from $p_{\theta}\left(x_{1:T}|y_{1:T}\right)$
and the expression for the marginal likelihood $p_{\theta}\left(y_{1:T}\right)$
are needed, their DPF approximation counterparts are used instead.

\lyxline{\normalsize}

\noindent \begin{center}
\textbf{PMMH sampler for SSSM} 
\par\end{center}

\hspace{-0.5cm}\underline{\textsf{Initialisation, }$i=0$}

$\hspace{-0.5cm}\bullet$ \textsf{Set }$\theta(0)$\textsf{ arbitrarily.}

$\hspace{-0.5cm}\bullet$ \textsf{Run the DPF targeting }$p_{\theta(0)}\left(x_{1:T}|y_{1:T}\right)$\textsf{,
sample }$X_{1:T}\left(0\right)\sim\widehat{p}_{\theta(0)}\left(\cdot|y_{1:T}\right)$
\textsf{and denote }$\widehat{p}_{\theta(0)}\left(y_{1:T}\right)$

$\hspace{-0.15cm}$\textsf{the marginal likelihood estimate.}

\hspace{-0.5cm}\underline{\textsf{For iteration }$i\geq1$}

$\hspace{-0.5cm}\bullet$ \textsf{Sample }$\theta^{\ast}\sim q\left(\cdot|\theta\left(i-1\right)\right)$\textsf{.}

$\hspace{-0.5cm}\bullet$ \textsf{Run the DPF targeting }$p_{\theta^{\ast}}\left(x_{1:T}|y_{1:T}\right)$\textsf{,
sample }$X_{1:T}^{\ast}\sim\widehat{p}_{\theta^{\ast}}\left(\cdot|y_{1:T}\right)$\textsf{
and denote }$\widehat{p}_{\theta^{\ast}}\left(y_{1:T}\right)$

$\hspace{-0.15cm}$\textsf{the marginal likelihood estimate.}

$\hspace{-0.5cm}\bullet$ \textsf{With probability} \begin{equation}
1\wedge\frac{\widehat{p}_{\theta^{\ast}}\left(y_{1:T}\right)\text{ }p\left(\theta^{\ast}\right)}{\widehat{p}_{\theta\left(i-1\right)}\left(y_{1:T}\right)\text{ }p\left(\theta\left(i-1\right)\right)}\frac{q\left(\theta\left(i-1\right)|\theta^{\ast}\right)}{q\left(\theta^{\ast}|\theta\left(i-1\right)\right)}\label{eq:acceptanceprobaPMMH}\end{equation}
 \textsf{\qquad{}\qquad{}\qquad{}\qquad{}}

\vspace{-0.25cm}
$\hspace{-0.15cm}$\textsf{set } $\theta\left(i\right)=\theta^{\ast}$\textsf{,
}$X_{1:T}\left(i\right)=X_{1:T}^{\ast}$\textsf{, }$\widehat{p}_{\theta\left(i\right)}\left(y_{1:T}\right)=\widehat{p}_{\theta^{\ast}}\left(y_{1:T}\right)$,

$\hspace{-0.15cm}$\textsf{otherwise set }$\theta\left(i\right)=\theta\left(i-1\right)$\textsf{,
}$X_{1:T}\left(i\right)=X_{1:T}\left(i-1\right)$\textsf{,}$\ \widehat{p}_{\theta\left(i\right)}\left(y_{1:T}\right)=\widehat{p}_{\theta\left(i-1\right)}\left(y_{1:T}\right).$

\lyxline{\normalsize}

\subsection{Particle Gibbs sampler\label{sec:particlegibbs_for_SSM}}

As discussed in Section \ref{sec:descriptionMCMC}, an attractive
but impractical strategy to sample from $p\left(\theta,x_{1:T}|y_{1:T}\right)$
consists of using the Gibbs sampler which iterates sampling steps
from $p_{\theta}\left(x_{1:T}|y_{1:T}\right)$ and $p\left(\theta|y_{1:T},x_{1:T}\right)$
or a modified Gibbs sampler where we insert a sampling step from $p_{\theta}\left(z_{0:T}|y_{1:T},x_{1:T}\right)$
after having sampled from $p_{\theta}\left(x_{1:T}|y_{1:T}\right)$
to update $\theta$ according to $p\left(\theta|y_{1:T},x_{1:T},z_{0:T}\right)$.
Numerous implementations rely on the fact that sampling from the conditional
density $p\left(\theta|y_{1:T},x_{1:T}\right)$ or $p\left(\theta|y_{1:T},x_{1:T},z_{0:T}\right)$
is feasible and thus the potentially difficult design of a proposal
density for $\theta$ can be bypassed. However, as mentioned before,
it is typically impossible to sample from $p_{\theta}\left(x_{1:T}|y_{1:T}\right)$.
Clearly substituting to the sampling step from $p_{\theta}\left(x_{1:T}|y_{1:T}\right)$,
sampling from the DPF approximation $\widehat{p}_{\theta}\left(x_{1:T}|y_{1:T}\right)$
would not provide Gibbs samplers admitting the correct invariant distribution.

We now present a valid particle approximation of the Gibbs sampler
which assumes we can sample from $p\left(\theta|y_{1:T},x_{1:T}\right)$.
Similarly it is possible to build a valid particle approximation of
the modified Gibbs sampler by the same arguments, but we omit the
details here for brevity.

\pagebreak{}

\lyxline{\normalsize}

\noindent \begin{center}
\textbf{PG sampler for SSSM}
\par\end{center}

\hspace{-0.5cm}\underline{\textsf{Initialisation, }$i=0$}

$\hspace{-0.5cm}\bullet$ \textsf{Set }$\theta\left(0\right),X_{1:T}\left(0\right)$\textsf{
arbitrarily.}

\hspace{-0.5cm}\underline{\textsf{For iteration }$i\geq1$}

$\hspace{-0.5cm}\bullet$ \textsf{Sample }$\theta\left(i\right)\sim p\left(\cdot|y_{1:T},X_{1:T}\left(i-1\right)\right)$\textsf{.}

$\hspace{-0.5cm}\bullet$ \textsf{Run a conditional DPF algorithm
targeting }$p_{\theta\left(i\right)}\left(x_{1:T}|y_{1:T}\right)$
\textsf{conditional upon} $X_{1:T}\left(i-1\right).$

$\hspace{-0.5cm}\bullet$ \textsf{Run a backward sampling algorithm
to obtain }$X_{1:T}\left(i\right)$\textsf{.}

\lyxline{\normalsize}

\noindent The remarkable property enjoyed by the PG\ algorithm is
that under weak assumptions it generates samples from $p\left(\theta,x_{1:T}|y_{1:T}\right)$
in steady state \emph{for any number} $N\geq2$ of particles used
to build the required DPF approximations. The non-standard steps of
the PG sampler are the conditional DPF algorithm and backward sampling
algorithms which we now describe.

Given a value of $\theta$ and a trajectory $x_{1:T}^{\ast}$, the
conditional DPF\ algorithm proceeds as follows.

\lyxline{\normalsize}

\noindent \begin{center}
\textbf{{Conditional DPF algorithm}} 
\par\end{center}

\hspace{-0.5cm}\underline{\textsf{At time }$n=1$}

$\hspace{-0.5cm}\bullet$ \textsf{Set } $\mathbf{S}_{1}=\mathcal{X}$
\textsf{ and for each }$x_{1}\in\mathcal{X}$ (which includes $x_{1}^{\ast}$),\textsf{
compute $m_{1|1}^{z,\theta}(x_{1})$ , $\Sigma_{1|1}^{z,\theta}(x_{1})$
and $g_{\theta}(y_{1}|x_{1})$ using the Kalman filter. }

$\hspace{-0.5cm}\bullet$\textsf{ Compute and normalise the weights.
For each $x_{1}\in\mathcal{X}$,}\begin{equation}
\overline{w}_{1}^{\theta}\left(x_{1}\right)=\nu_{\theta}\left(x_{1}\right)g_{\theta}\left(\left.y_{1}\right\vert x_{1}\right),\text{ }W_{1}^{\theta}\left(x_{1}\right)=\frac{\overline{w}_{1}^{\theta}\left(x_{1}\right)}{{\textstyle \sum\nolimits _{x_{1}^{\prime}\in\mathcal{X}}}\overline{w}_{1}^{\theta}\left(x_{1}^{\prime}\right)}.\end{equation}

\vspace{-0.25cm}
\hspace{-0.5cm}\underline{\textsf{At times
}$n=2,...,T$}

$\hspace{-0.5cm}\bullet$ \textsf{If $\left|\mathbf{S}_{n-1}\right|\leq N$
set $C_{n-1}=\infty$ otherwise set }$C_{n-1}$\textsf{ to the unique
solution of} \[
\sum_{x_{1:n-1}\in\mathbf{S}_{n-1}}1\wedge C_{n-1}W_{n-1}^{\theta}\left(x_{1:n-1}\right)=N.\]

$\hspace{-0.5cm}\bullet$ \textsf{If }$W_{n-1}^{\theta}\left(x_{1:n-1}^{\ast}\right)>1/C_{n-1}$\textsf{,
maintain the }$L_{n-1}$\textsf{ trajectories which have weights strictly
superior to }$1/C_{n-1}$\textsf{ (which includes }$x_{1:n-1}^{\ast}$\textsf{),
then apply the stratified resampling mechanism to the other }$N\left|\mathcal{X}\right|-L_{n-1}$\textsf{
weighted trajectories to yield $N-L_{n-1}$ survivors. Set} $\mathbf{S}_{n-1}'$
\textsf{ to the set of surviving and maintained trajectories.}

$\hspace{-0.5cm}\bullet$ \textsf{If }$W_{n-1}^{\theta}\left(x_{1:n-1}^{\ast}\right)\leq1/C_{n-1}$
\textsf{maintain the }$L_{n-1}$\textsf{ trajectories which have weights
strictly superior to }$1/C_{n-1}$\textsf{ (which excludes }$x_{1:n-1}^{\ast}$\textsf{),
then apply the conditional stratified resampling mechanism to the
other }$N\left|\mathcal{X}\right|-L_{n-1}$\textsf{ weighted trajectories
to yield $N-L_{n-1}$ survivors (which include $x_{1:n}^{*}$). Set}
$\mathbf{S}_{n-1}'$ \textsf{ to the set of surviving and maintained
trajectories.}

$\hspace{-0.5cm}\bullet$ \textsf{Set} $\mathbf{S}_{n}=\mathbf{S}_{n-1}'\times\mathcal{X}$.

$\hspace{-0.5cm}\bullet$ \textsf{For each $x_{1:n}\in\mathbf{S}_{n}$,
update and store $m_{n|n}^{z,\theta}(x_{1:n})$ and $\Sigma_{n|n}^{z,\theta}(x_{1:n})$
and compute $g_{\theta}(y_{n}|y_{1:n-1},x_{1:n})$ using the Kalman
filter.}

$\hspace{-0.5cm}\bullet$\textsf{ Compute and normalise the weights.
For each $x_{1:n}\in\mathcal{X}^{n}$,}

\begin{align}
\overline{w}_{n}^{\theta}\left(x_{1:n}\right) & =f_{\theta}(x_{n}|x_{1:n-1})g_{\theta}(y_{n}|y_{1:n-1},x_{1:n})\dfrac{W_{n-1}^{\theta}\left(x_{1:n-1}\right)}{1\wedge C_{n-1}W_{n-1}^{\theta}\left(x_{1:n-1}\right)}\label{eq:updateweight1_cond}\\
W_{n}^{\theta}\left(x_{1:n}\right) & =\frac{\overline{w}_{n}^{\theta}\left(x_{1:n}\right)}{{\textstyle \sum\nolimits _{x_{1:n}^{\prime}\in\mathbf{S}_{n}}}\overline{w}_{n}^{\theta}\left(x_{1:n}^{\prime}\right)}.\label{eq:updateweight2_cond}\end{align}

$\hspace{-0.5cm}\bullet$ \textsf{If backward sampling is to be used,
store $W_{n}^{\theta}(x_{1:n})$, $m_{n|n}^{z,\theta}(x_{1:n})$ and
$\Sigma_{n|n}^{z,\theta}(x_{1:n})$ for each $x_{1:n}\in\mathbf{S}_{n}$.}

\lyxline{\normalsize}

\noindent The conditional stratified resampling procedure can be implemented
as follows.

\pagebreak{}

\noindent \lyxline{\normalsize}

\noindent \begin{center}
\textbf{Conditional stratified resampling} 
\par\end{center}

$\hspace{-0.5cm}\bullet$ \textsf{Normalise the weights }$\overline{w}_{n-1}^{\theta}\left(x_{1:n-1}\right)$\textsf{
of the }$N\left|\mathcal{X}\right|-L_{n-1}$\textsf{ particles and
label them according to the order of the corresponding $x_{1:n-1}$
to obtain }$\widehat{W}_{n-1}^{\theta}\left(x_{1:n-1}^{\left(i\right)}\right)$
\textsf{;} $i=1,...,N\left|\mathcal{X}\right|-L_{n-1}$. \textsf{Define}
$\kappa$ \textsf{to be the integer satisfying} $x_{1:n-1}^{(\kappa)}=x_{1:n-1}^{*}.$

$\hspace{-0.5cm}\bullet$\textsf{ Construct the corresponding cumulative
distribution function: for} $i=1,...,N\left|\mathcal{X}\right|-L_{n-1}$,

\[
Q_{n-1}^{\theta}(i):=\sum_{j\leq i}\widehat{W}_{n-1}^{\theta}\left(x_{1:n-1}^{\left(j\right)}\right),\qquad Q_{n-1}^{\theta}(0):=0.\]

$\hspace{-0.5cm}\bullet$ \textsf{Sample }$U_{*}$\textsf{ uniformly
on }$\left[Q_{n-1}^{\theta}(\kappa-1),Q_{n-1}^{\theta}(\kappa)\right]$,
\textsf{set} $U_{1}=U_{*}-\dfrac{\left\lfloor \left(N-L_{n-1}\right)U_{*}\right\rfloor }{N-L_{n-1}}$\textsf{
and compute }$U_{j}=U_{1}+\frac{j-1}{N-L_{n-1}}$\textsf{ for }$j=2,...,N-L_{n-1}$.
\textsf{Here }$\left\lfloor a\right\rfloor $ \textsf{denotes the
largest integer not greater than} $a$.

$\hspace{-0.5cm}\bullet$ \textsf{For} $i=1,...,N\left|\mathcal{X}\right|-L_{n-1}$,
\textsf{if there exists} $j\in\left\{ 1,...,N-L_{n-1}\right\} $ \textsf{such
that} $Q_{n-1}^{\theta}(i-1)<U_{j}\leq Q_{n-1}^{\theta}(i)$, \textsf{then}
$x_{1:n-1}^{(i)}$ \textsf{survives}.

\lyxline{\normalsize}

\noindent The backward sampling step is an important component of
the PG algorithm. In contrast to the standard PMCMC algorithms of
\citet{andrieudoucetholenstein2008}, it allows the sampled trajectory
obtained from the conditional SMC update not only to be chosen from
those surviving at time $T$, but allows full exploration of all trajectories
sampled during the Conditional DPF algorithm. Further comments on
the theoretical validity of alternative schemes are made in section
\ref{sec:proofs} and demonstration of numerical performance given
in section \ref{sec:applications}.

We note that this procedure is of some independent interest for smoothing
in SSSM's if $\theta$ is known, as it can be combined with the standard
DPF algorithm. A forward filtering-backward smoothing algorithm for
SSSM was devised in \citet{Fong2002}, and involved joint sampling
of both continuous and discrete variables from an approximation of
$p_{\theta}(x_{1:T},z_{0:T}|y_{1:T})$. The backward sampling algorithm
we propose is different because the continuous component of the state
is integrated out analytically, giving a further Rao-Blackwellization
over the scheme of \citet{Fong2002}. Furthermore, the fact that the
backward sampling algorithm involves sampling only discrete--valued
variables is central to the validity of the PG algorithm, discussed
in the next section. Details of the matrix-vector recursions necessary
for the implementation of the backward sampling procedure are given
in Appendix \ref{sec:Backward-Sampling}.

\lyxline{\normalsize}

\noindent \begin{center}
\textbf{Backward Sampling} 
\par\end{center}

\hspace{-0.5cm}\underline{\textsf{At time }$n=T$}

$\hspace{-0.5cm}\bullet$ \textsf{Sample a path }$X_{1:T}^{*}$\textsf{
from the distribution on }$\mathbf{S}_{T}\subset\mathcal{X}^{T}$\textsf{
defined by }$\{W_{T}^{\theta}(x_{1:T})\}$, \textsf{ then discard
}$X_{1:T-1}^{*}$\textsf{ to yield }$X_{T}'=X_{T}^{*}$. Set $\Xi_{T}=0$,
$\mu_{T}=0$.

\hspace{-0.5cm}\underline{\textsf{At times }$n=T-1,...,1$}

$\hspace{-0.5cm}\bullet$ \textsf{Update }$\Xi_{n}$ and $\mu_{n}$
\textsf{as per the procedure of Appendix }\ref{sec:Backward-Sampling}.

$\hspace{-0.5cm}\bullet$ \textsf{For each $x_{1:n}\in\mathbf{S}_{n}$}
\textsf{compute the backward weight}

\[
V_{n}^{\theta}\left(x_{1:n}\left|x_{n+1:T}'\right.\right)\propto W_{n}^{\theta}(x_{1:n})p_{\theta}(x_{n+1:T}'|x_{1:n})p_{\theta}(y_{n+1:T}|y_{1:n},x_{1:n},x_{n+1:T}')\]
 \textsf{where} \[
p_{\theta}(y_{n+1:T}|y_{1:n},x_{1:n},x_{n+1:T}')=\int p_{\theta}(y_{n+1:T}|z_{n},x_{n+1:T}')p_{\theta}(z_{n}|x_{1:n},y_{1:n})dz_{n}\]
 \textsf{is evaluated using $\mu_{n}$, $\Xi_{n}$ and the stored
}$m_{n|n}^{z,\theta}(x_{1:n})$\textsf{ and} $\Sigma_{n|n}^{z,\theta}(x_{1:n})$\textsf{
of} $p_{\theta}(z_{n}|x_{1:n},y_{1:n})$ \textsf{as per Eq. \eqref{eq:backward_weight_like}
in Appendix \ref{sec:Backward-Sampling}.}

$\hspace{-0.5cm}\bullet$\textsf{ Normalise the backward weights $\left\{ V_{n}^{\theta}\left(x_{1:n}\left|x_{n+1:T}'\right.\right)\right\} _{x_{1:n}\in\mathbf{S}_{n}}$
and draw from the distribution they define on} $\mathbf{S}_{n}\subset\mathcal{X}^{n}$
\textsf{to obtain} $X_{1:n}'$.

$\hspace{-0.5cm}\bullet$ \textsf{If } $n>1$ \textsf{discard} $X_{1:n-1}'$,
\textsf{ otherwise output } $X_{1:T}'$.

\lyxline{\normalsize}

\subsection{Validity of the algorithms\label{sec:proofs}}

\noindent The key to establishing the validity of the PMCMC algorithms
is in showing that these are standard MCMC algorithms on an extended
state-space including all the random variables introduced in the DPF
algorithm. 

The first step is to observe that under our representation of the
DPF algorithm, its operation remains essentially unchanged if at each
iteration we adopt the convention of setting $\overline{w}_{n}^{\theta}(x_{1:n})=W_{n}^{\theta}\left(x_{1:n}\right)=0$
for all $x_{1:n}\notin\mathbf{S}_{n}$, and to replace all summations
over $\mathbf{S}_{n}$ with summations over $\mathcal{X}^{n}$. We
assume this convention throughout the remainder of this section, i.e.
from now on, $\mathbf{W}_{n}^{\theta}$ is a set of weights over $\mathcal{X}^{n}$,
but only those weights over $\mathbf{S}_{n}$ are non-zero. In this
case the solution of $\sum_{x_{1:n}\in\mathbf{S}_{n}}1\wedge C_{n}W_{n}^{\theta}\left(x_{1:n}\right)=N$
is identical to the solution of $\sum_{x_{1:n}\in\mathcal{X}^{n}}1\wedge C_{n}W_{n}^{\theta}\left(x_{1:n}\right)=N$.
Furthermore we can consider the resampling mechanism as acting on
all trajectories in $\mathcal{X}^{n}$ and not only those in $\mathbf{S}_{n}$;
those with zero weights clearly fall below the threshold $1/C_{n}$
and there is zero probability of them surviving the resampling operation.
As we shall see, the intuitive implication of this observation is
that once a trajectory $x_{1:n}$ has been discarded, it is lost and
for any $m>n$ , any subsequent trajectory $(x_{1:n},x_{n+1:m}')\in\mathcal{X}^{m}$
is also assigned zero weight. We denote by $\mathbf{W}_{n}^{\theta}$
the set of normalised importance weights at time $n$, that is $\mathbf{W}_{n-1}^{\theta}:=\{W_{n-1}^{\theta}(x_{1:n-1}),x_{1:n-1}\in\mathcal{X}^{n-1}\}$

We next write an expression for the joint distribution of the sequence
of random support sets $\mathbf{S}_{1},\mathbf{S}_{2},...,\mathbf{S}_{T}$
generated through the DPF algorithm. By definition of the algorithm,
for $n\geq2$, $\mathbf{S}_{n}$ is conditionally independent of the
history of the algorithm given $\mathbf{W}_{n-1}^{\theta}$ \begin{equation}
\mathbf{S}_{n}|\left(\mathbf{W}_{n-1}^{\theta}=\mathbf{w}_{n-1}^{\theta}\right)\sim r_{n}^{N}(\cdot|\mathbf{w}_{n-1}^{\theta}),\label{eq:distributionsurvival}\end{equation}
where for each $N$, $n$ and $\mathbf{w}_{n-1}^{\theta}$, $r_{n}^{N}(\cdot|\mathbf{w}_{n-1}^{\theta})$
can be understood as a probability distribution over the set of subsets
of $\mathcal{X}^{n}$, and we denote this set of subsets by $\mathcal{P}(\mathcal{X}^{n})$.
This distribution is parameterized by $N$ because for all $n\geq2$,
for each point $\mathbf{s}_{n}$ in the support of $r_{n}^{N}(\cdot|\mathbf{w}_{n-1}^{\theta})$
, $|\mathbf{s}_{n}|=N|\mathcal{X}|$ . In the case of $n=1$, $r_{n}^{N}(\cdot)=\mathbb{I}[\cdot=\mathcal{X}]$.

We will not need an explicit expression for the distribution (\ref{eq:distributionsurvival}),
but from the definition of the optimal resampling mechanism \citep{fearnhead1998,fearnhead2003},
we know that it has the following marginal property: for all $x_{1:n}\in\mathcal{X}^{n}$
, we have\begin{equation}
r_{n}^{N}(x_{1:n}\in\mathbf{s}_{n}|\mathbf{w}_{n-1}^{\theta})=1\wedge C_{n-1}w_{n-1}^{\theta}\left(x_{1:n-1}\right).\label{eq:propertymarginalresampling}\end{equation}
where we have adopted the abusive notation that \[
r_{n}^{N}(x_{1:n}\in\mathbf{s}_{n}|\mathbf{w}_{n-1}^{\theta}):=\sum_{\mathbf{s}_{n}':x_{1:n}\in\mathbf{s}_{n}'}r_{n}^{N}(\mathbf{s}_{n}'|\mathbf{w}_{n-1}^{\theta})\]
Eq. (\ref{eq:propertymarginalresampling}) implies \[
r_{n}^{N}(x_{1:n}\in\mathbf{s}_{n}|w_{n-1}^{\theta}\left(x_{1:n-1}\right)=0)=0.\]
Combined with Eq. (\ref{eq:distributionsurvival}) we see that for
any $n$ and $x_{1:n-1}$, conditional on the event that $W_{n-1}^{\theta}\left(x_{1:n-1}\right)=0$,
any subsequent paths which have $x_{1:n-1}$ as their first $n-1$
coordinates are also assigned zero weight and are not members of any
subsequent $\mathbf{S}_{n}$. Thus the corresponding subsequent weights
need never be computed or stored, as required to control the cost
of the algorithm. We thus have the property as claimed earlier that
once a trajectory is discarded it is not recovered. To summarize the
law of the DPF algorithm, we can write the density of $\mathbf{S}_{1},\mathbf{S}_{2},...,\mathbf{S}_{T}$
on $\prod_{n=1}^{T}\mathcal{P}(\mathcal{X}^{n})$ as\begin{equation}
\psi_{\theta}^{N}(\mathbf{s}_{1},\mathbf{s}_{2},...,\mathbf{s}_{T})=r_{1}^{N}(\mathbf{s}_{1})\prod_{n=2}^{T}r_{n}^{N}(\mathbf{s}_{n}|\mathbf{w}_{n-1}^{\theta}).\label{eq:DPFsamplingdistribution}\end{equation}
 As the weights $\mathbf{W}_{n}^{\theta}$ are just a deterministic
function of $\mathbf{S}_{1},\ldots,\mathbf{S}_{n}$, it is not necessary
to introduce them as arguments of $\psi_{\theta}^{N}$.

The key to the PMCMC algorithms described here is to define the following
artificial target density on $\Theta\times\mathcal{X}^{T}\times\prod_{n=1}^{T-1}\mathcal{P}(\mathcal{X}^{n})$
through \begin{equation}
\pi^{N}(\theta,x_{1:T},\mathbf{s}_{1},\mathbf{s}_{2},...,\mathbf{s}_{T})=p(\left.\theta,x_{1:T}\right\vert y_{1:T})\left\{ \prod_{n=2}^{T}\mathbb{I}[x_{1:n}\in\mathbf{s}_{n}]\right\} \frac{\psi_{\theta}^{N}(\mathbf{s}_{1},\mathbf{s}_{2},...,\mathbf{s}_{T})}{\prod_{n=2}^{T}r_{n}^{N}(x_{1:n}\in\mathbf{s}_{n}|\mathbf{w}_{n-1}^{\theta})}\label{eq:artificialdistributiondetails}\end{equation}
which admits $p(\left.\theta,x_{1:T}\right\vert y_{1:T})$ as a marginal
by construction. Let $\pi_{\theta}^{N}(x_{1:T},\mathbf{s}_{1},\mathbf{s}_{2},...,\mathbf{s}_{T})$
denote the density of $X_{1:T},\mathbf{S}_{1},\mathbf{S}_{2},...,\mathbf{S}_{T}$
conditional upon $\theta$ under $\pi^{N}(\theta,x_{1:T},\mathbf{s}_{1},\mathbf{s}_{2},...,\mathbf{s}_{T})$.
In the following results we show that the PMMH and PG algorithms are
just standard MCMC updates targeting this artificial distribution.
Proofs can be found in Appendix \ref{app:Proofs}.

We first present a result establishing the convergence of the PMMH
sampler which relies on the following assumption.

\begin{condition} \label{hyp:trueMHconverges}The MH sampler of target
density $p\left(\left.\theta\right\vert y_{1:T}\right)$ and proposal
density $q(\theta^{\ast}|\theta)$ is irreducible and aperiodic (and
hence converges for almost all starting points). 

\end{condition}

We have the following result.

\begin{theorem} \label{theorem:particleMMH}For any $N\geq2$
\begin{enumerate}
\item the PMMH sampler is an MH\ sampler defined on the extended space
$\Theta\times\mathcal{X}^{T}\times\prod_{n=1}^{T}\mathcal{P}(\mathcal{X}^{n})$
with target density $\pi^{N}(\theta,x_{1:T},\mathbf{s}_{1},\mathbf{s}_{2},...,\mathbf{s}_{T})$
defined in Eq. (\ref{eq:artificialdistributiondetails}) and proposal
density \begin{equation}
q(\theta^{\ast}|\theta)\text{ }w_{T}^{\theta^{\ast}}\left(x_{1:T}^{\ast}\right)\text{ }\psi_{\theta^{\ast}}^{N}\left(\mathbf{s}_{1}^{\ast},\mathbf{s}_{2}^{\ast},...,\mathbf{s}_{T}^{\ast}\right)\label{eq:proposalPMMH}\end{equation}
 where $w_{T}^{\theta^{\ast}}\left(x_{1:T}^{\ast}\right)$ is the
realisation of the normalised importance weight associated to the
population of particles proposed by the DPF\ algorithm.
\item if additionally (A\ref{hyp:trueMHconverges}) holds, the PMMH sampler
generates a sequence $\left\{ \theta\left(i\right),X_{1:T}\left(i\right)\right\} $
whose marginal distributions $\{\mathcal{L}^{N}\left(\left(\theta\left(i\right),X_{1:T}\left(i\right)\right)\in\cdot\right)\}$
satisfy \[
\left\Vert \mathcal{L}^{N}\left(\left(\theta\left(i\right),X_{1:T}\left(i\right)\right)\in\cdot\right)-p\left(\left.\cdot,\cdot\right\vert y_{1:T}\right)\right\Vert _{TV}\rightarrow0\ \text{as}\ i\rightarrow\infty\ .\]
for almost all starting points. \bigskip{}
 
\end{enumerate}
\end{theorem}

Next we consider the backward sampling procedure and establish its
invariance properties.

\begin{proposition}\label{prop:backward_sampling} For any $N\geq2$
and $\theta\in\Theta$, assume $(X_{1:T},\mathbf{S}_{1},\mathbf{S}_{2},...,\mathbf{S}_{T})$
is distributed according to $\pi_{\theta}^{N}(\cdot)$ and let $X_{1:T}'$
be the trajectory obtained at \emph{any} time step $m$ of the backward
sampling procedure operating on $(X_{1:T},\mathbf{S}_{1},\mathbf{S}_{2},...,\mathbf{S}_{T})$.
Then $X_{1:T}'$ is distributed according to $p_{\theta}\left(x_{1:T}|y_{1:T}\right)$. 

\end{proposition}

We now state a sufficient condition for the convergence of the PG
sampler and provide a simple convergence result.

\begin{condition} \label{hyp:Gibbsconverges} The Gibbs sampler defined
by drawing alternately from the conditionals $p\left(\theta|y_{1:T},x_{1:T}\right)$
and $p_{\theta}\left(x_{1:T}|y_{1:T}\right)$ is irreducible and aperiodic
(and hence converges for $p$-almost all starting points). \end{condition}

We have the following result.

\begin{theorem} \label{theorem:particlegibbs}$\;$
\begin{enumerate}
\item steps $1-4$ of the PG update define a transition kernel on the extended
space $\Theta\times\mathcal{X}^{T}\times\prod_{n=1}^{T}\mathcal{P}(\mathcal{X}^{n})$
of invariant density $\pi^{N}(\theta,x_{1:T},\mathbf{s}_{1},\mathbf{s}_{2},...,\mathbf{s}_{T})$
defined in Eq. (\ref{eq:artificialdistributiondetails}) for any $N\geq2$.
\item if additionally (A\ref{hyp:Gibbsconverges}) holds, the PG sampler
generates a sequence $\left\{ \theta\left(i\right),X_{1:T}\left(i\right)\right\} $
whose marginal distributions $\{\mathcal{L}_{PG}^{N}\left(\left(\theta\left(i\right),X_{1:T}\left(i\right)\right)\in\cdot\right)\}$
satisfy for any $N\geq2$ \[
\left\Vert \mathcal{L}_{PG}^{N}\left(\left(\theta\left(i\right),X_{1:T}\left(i\right)\right)\in\cdot\right)-p\left(\left.\cdot,\cdot\right\vert y_{1:T}\right)\right\Vert _{tv}\rightarrow0\ \text{as}\ i\rightarrow\infty,\]
 for almost all starting points. 
\end{enumerate}
\end{theorem}

\begin{remark}\label{rem:backward_sampling}

The reader will observe that as Proposition \ref{prop:backward_sampling}
applies for any time step of the backward sampling, modification of
the PG algorithm to the case where $X_{1:T}\left(i\right)$ is set
to the $X_{1:T}'$ obtained at any time step of the backward sampling
procedure also corresponds to a Markov kernel of the required invariant
distribution. For example, one could simply apply only the first backward
sampling step: sample $X_{1:T}'$ from the distribution defined by
$\left\{ W_{T}(x_{1:T})\right\} $ and then set $X_{1:T}\left(i\right)=X_{1:T}'$.
The resulting algorithm is closer akin to the original Particle Gibbs
algorithm of \citet{andrieudoucetholenstein2008}. However, in numerical
experiments in the context of SSSMs this approach has been found to
be relatively inefficient. This phenomenon is discussed further and
demonstrated numerically in section \ref{sec:applications}. 

\end{remark}

\section{Applications\label{sec:applications}}

\subsection{Example 1: Autoregression with shifting level}

In our first numerical experiments we return to the toy model specified
in section \ref{subsec:Example:-Auto-regression-with} and address
some generic issues regarding algorithmic settings and performance.

\subsubsection{Particle Gibbs and the effect of backward sampling}

We first demonstrate the effect of applying the backward sampling
procedure as part of the PG algorithm. The purpose of this section
is to show the importance of applying backward sampling as part of
the PG algorithm and to show its advantage over the standard Gibbs
sampler. From hereon we refer to as ``PG without backward sampling''
the alternative PG scheme described in Remark \ref{rem:backward_sampling}
which involves sampling $X_{1:T}'$ from the distribution defined
by $\left\{ W_{T}(x_{1:T})\right\} $ and immediately setting $X_{1:T}\left(i\right)=X_{1:T}'$. 

Recall that for this model the parameters are $\theta=[\phi\;\sigma^{2}\; P_{X}]$.
Conjugate priors are readily available: a Gaussian distribution for
$\phi$, inverse-gamma for $\sigma^{2}$ and independent Dirichlet
for each row of $P_{X}$. A data record of length $T=1000$ was generated
from the model with true parameter values of $\phi=0.1$, $\sigma=0.1$
and $P_{X}=\left[\begin{array}{cc}
0.99 & 0.01\\
0.99 & 0.01\end{array}\right]$. Flat Dirichlet priors were set on each row of $P_{X}$. A $\mathcal{N}(0,10)$
distribution restricted to $\left|\phi\right|\leq1$ was set over
$\phi$ and a $(0.1,0.1)$ inverse gamma distribution was set over
$\sigma^{2}$. The initial distribution over $\mu_{0}$ was $\mathcal{N}(0,10)$.
For various numbers of particles the PG algorithm, with and without
backward sampling, was run and compared to the standard one-at-a-time
Gibbs algorithm in terms of the sample lag $1$ autocorrelation for
each component of the discrete latent trajectory $\left\{ X_{n}\right\} $.
In all cases the simulation smoother of \citet{durbin2002} was used
to sample from $p_{\theta}(z_{0:T}|y_{1:T},x_{1:T})$.

\begin{figure}
\begin{minipage}[t]{0.48\columnwidth}%
\begin{center}
\includegraphics[width=1\textwidth]{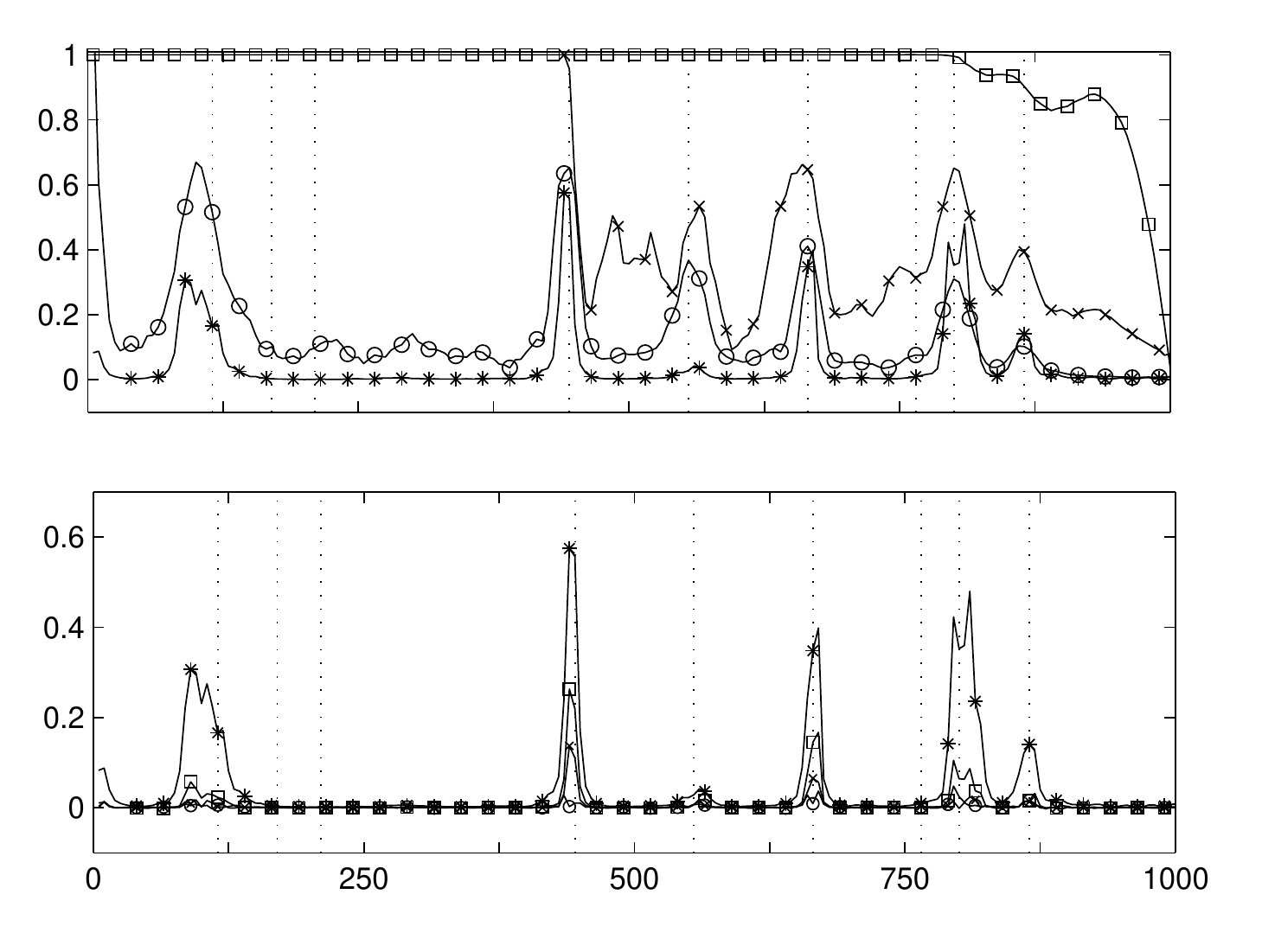}\caption{Example 1. Sample lag-$1$ autocorrelation for each of the discrete
trajectory components $\left\{ X_{n}(i),\: n=1,...,1000\right\} $
with (bottom) and without (top) backward sampling for various numbers
of particles: $\square$: $N=10$; $\times$: $N=20$; $\bigcirc$:
$N=50$; In both top and bottom $*$ is sample autocorrelation for
standard one-at-a-time Gibbs. Vertical dashed lines are true locations
of $X_{n}=1$.}
\label{Flo:fig:toy_model_pg_ac}
\par\end{center}%
\end{minipage}\hfill{}%
\begin{minipage}[t]{0.48\columnwidth}%
\begin{center}
\includegraphics[width=1\textwidth]{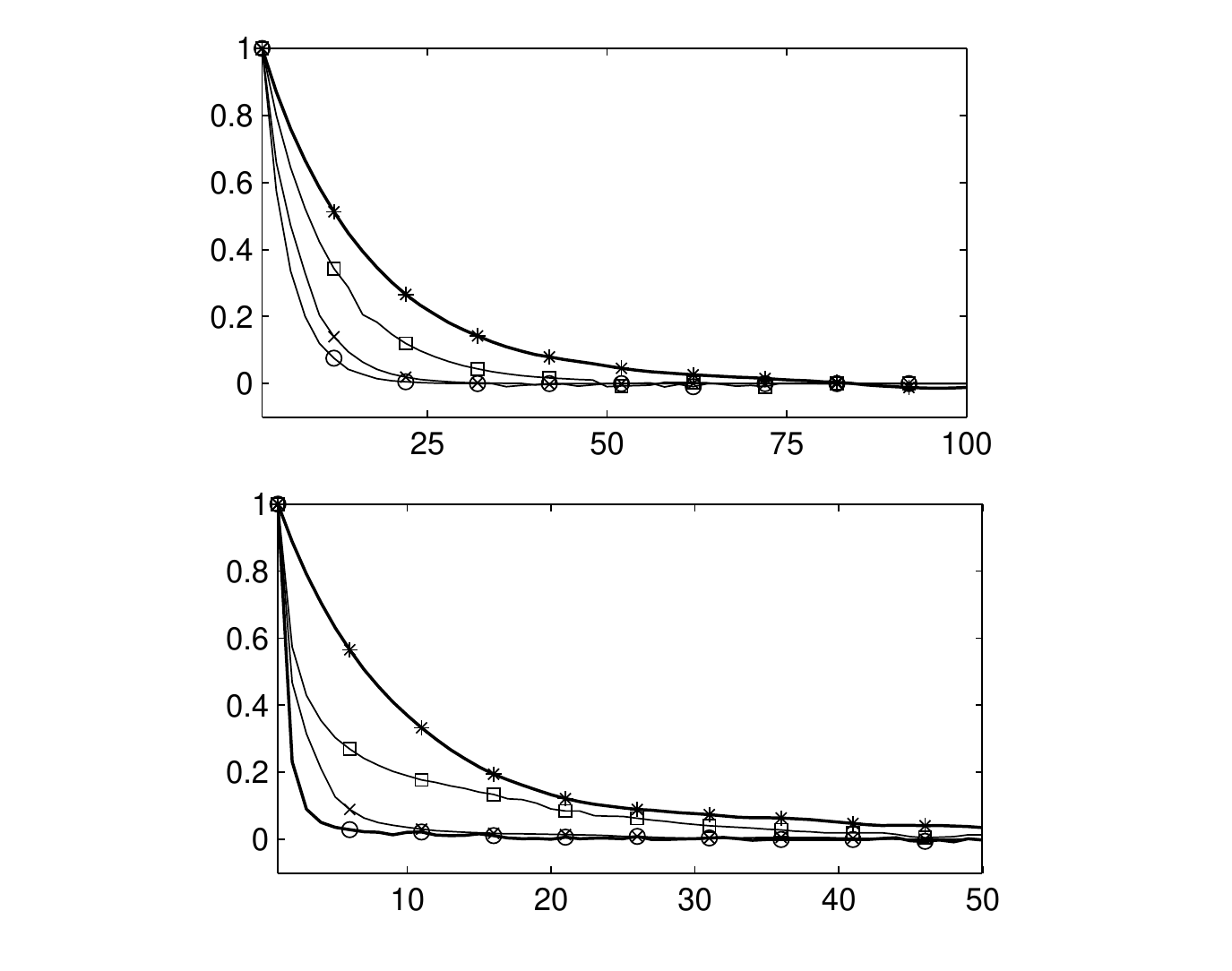}\caption{Example 1. Autocorrelation against lag for standard Gibbs sampler
($*$) and PG with backward sampling and various numbers of particles:
$\square$: $N=10$; $\times$: $N=20$; $\bigcirc$: $N=50$. Top
pane is for $\phi$ and bottom pane for $\sigma^{2}$. }
\label{Flo:fig2}
\par\end{center}%
\end{minipage}
\end{figure}

In both panes of Figure \ref{Flo:fig:toy_model_pg_ac}. the vertical
dashed lines show the true times at which $X_{n}=1$. The bottom pane
shows the lag $1$ autocorrelation for PG with backward sampling and
the standard one-at-a-time Gibbs sampler: here it was found that for
all components of the trajectory, increasing $N$ monotonically decreased
the autocorrelation and for any $N$ the PG algorithm exhibited lower
autocorrelation than the standard one-at-a-time algorithm. Spikes
in the autocorrelation coincide with the true times at which $X_{n}=1$
and between these times the autocorrelation, even using the standard
Gibbs sampler, was found to be very low. By contrast, for the PG without
backward sampling and the same numbers of particles, the autocorrelation
from the PG algorithm was higher than that from the standard Gibbs
algorithm for most components of the discrete trajectory. In all cases
the sample autocorrelation was computed from $10^{5}$ iterations
after a burn-in of $10^{4}$ iterations. After the $10^{4}+10^{5}$
iterations, with $N=10$ and $N=20$ particles, the PG without backward
sampling had entirely failed to converge: in the plots of Figure \ref{Flo:fig:toy_model_pg_ac},
we use the ranges in which the plots reach the value exactly $1$
to represent those components of the discrete trajectory never having
changed from their initial condition (such a sample sequence does
not have a well defined autocorrelation as its sample variance is
zero). Very similar results were observed for other initialisations
and data records. 

This performance can be explained in terms of the well-known particle
path degeneracy phenomenon which arises from the resampling mechanism
in SMC algorithms: the act of repeated selection of sampled paths
inevitably leads to a loss in diversity in their early components.
In the present context the path degeneracy influences the performance
of the PG algorithms via the conditional DPF update. During the conditional
DPF operation at MCMC iteration $i+1$, by construction of the conditional
DPF, $X_{1:T}\left(i\right)$ is forced to survive until time step
$T$. Thus for the PG without backward sampling, for some $m<T$,
the path degeneracy phenomenon implies there is a significant probability
that $X_{1:m}\left(i\right)$ coincides with $X_{1:m}(i+1)$. This
explains the strong correlations between components of consecutive
samples of the latent trajectory shown in the top pane of Figure \ref{Flo:fig:toy_model_pg_ac}.
By contrast, backward sampling provides a chance for the path degeneracy
to be circumvented. The CPU time for one iteration of the PG with
backward sampling was found to be between $1$ and $1.5$ times that
without backward sampling for the same number of particles. The results
therefore indicate that overall it is significantly more efficient
to use the backward sampling method and from now on it is the only
PG algorithm we consider.

Figure \ref{Flo:fig2} shows sample autocorrelation as a function
of lag for various numbers of particles from the PG algorithm with
backward sampling and the standard one-at-a-time Gibbs sampler. We
observe that using large $N$ leads to lower autocorrelation and very
little decrease in autocorrelation was observed using more than $N=50$
particles. As we go on to discuss in more details in the next section,
under the Dirichlet prior for each row of $P_{X}$ it is possible
to analytically integrate out $P_{X}$ both when using the standard
Gibbs sampler and the PG, and we did so. The above experiments were
also conducted in the case where $P_{X}$ is not integrated out and
we obtained results which were almost identical (not shown). 

\subsubsection{Treatment of $P_{X}$}

A common feature of SSSMs is that it is possible to analytically integrate
out $P_{X}$ under Dirichlet priors for each of its rows and the autoregressive
model with shifting level is no exception. It is natural to ask, even
in the context of standard MCMC algorithms, whether it is beneficial
to perform this integration analytically, or to treat $P_{X}$ as
part of the sampling problem. To the authors' knowledge, in the context
of SSSMs this issue has not been treated in the literature.

Consider first the standard one-at-a-time Gibbs sampling case. The
reader will recall from section \ref{sec:descriptionMCMC} and \citet{gerlach2000}
that the algorithm involves sampling from \begin{equation}
p_{\theta}\left(x_{n}|y_{1:T},x_{1:n-1},x_{n+1:T}\right)\propto p_{\theta}(y_{n}|y_{1:n-1},x_{1:n})p_{\theta}(x_{n}|x_{1:n-1},x_{n+1:T})p_{\theta}(y_{n+1:T}|y_{1:n},x_{1:T})\label{eq:ex1:full_conditional}\end{equation}
for each $n$. Conditionally on $P_{X}$, the process $\left\{ X_{n}\right\} _{n\geq1}$
is Markov and so in the above display we have the simplification $p_{\theta}(x_{n}|x_{1:n-1},x_{n+1:T})=p_{\theta}(x_{n}|x_{n-1},x_{n+1})$.
Conversely, when $P_{X}$ is integrated out, in which case the parameter
reduces to $\theta=[\phi\;\sigma^{2}]$, the process $\left\{ X_{n}\right\} _{n\geq1}$
is not Markov and the former simplification is not applicable. Thus,
in terms of the correlation structure of the Markov chains generated
by the corresponding Gibbs samplers, there appears to be a trade-off
between conditioning on $P_{X}$ and conditioning on components of
the $\left\{ X_{n}\right\} _{n\geq1}$ process when drawing from distributions
of the form (\ref{eq:ex1:full_conditional}). In terms of computational
cost there is no significant difference: in the case that $P_{X}$
is integrated out analytically evaluation of $p_{\theta}(x_{n}|x_{1:n-1},x_{n+1:T})$
requires only state-transition count statistics which are cheap to
compute and store. 

Analogous remarks to those above hold for the PG algorithm. It involves
computing $f_{\theta}(x_{n}|x_{1:n-1})$ in the conditional DPF step
and $p_{\theta}(x_{n+1:T}|x_{1:n})$ in the backward sampling step
and it is in these places that the same conditioning issues arise.
In our numerical experiments for this model and others we were unable
to establish that either incorporating $P_{X}$ into the sampling
problem or integrating it out analytically lead to a significant advantage
in terms of sample autocorrelation, both for the standard one-at-a-time
Gibbs sampler and the PG algorithm (results not shown). It would be
very interesting to study the theoretical properties underlying this
issue in Gibbs sampling algorithms for SSSMs but such an investigation
is well beyond the scope of this article.

\begin{figure}
\begin{minipage}[t]{0.49\columnwidth}%
\begin{center}
\includegraphics[width=1\textwidth]{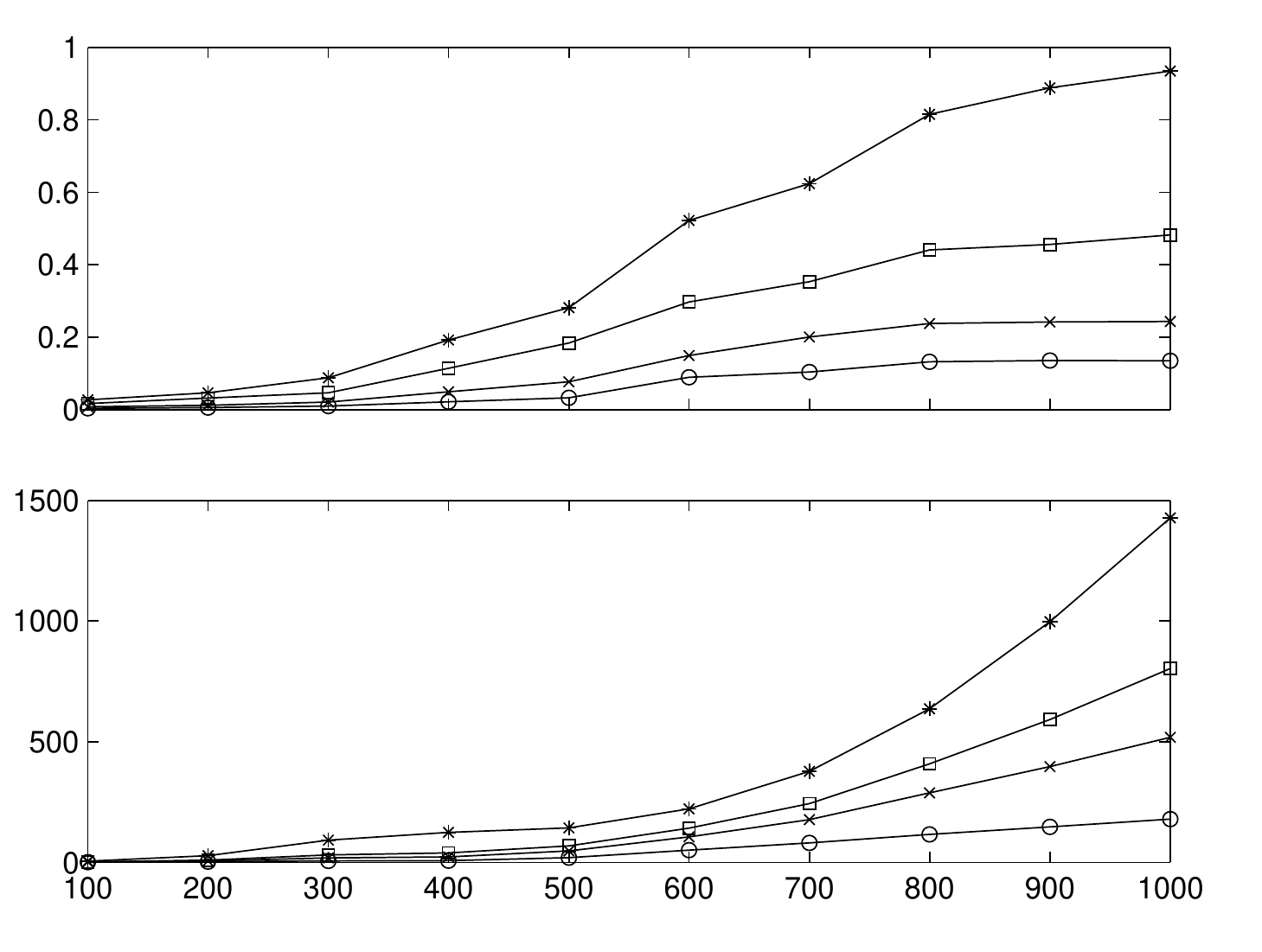}\caption{Example 1. Sample variance of $\log\widehat{p}_{\theta}(y_{1:n})$
for fixed $\theta$ as a function of $n$ for various numbers of particles:
$*$: $N=10$; $\square$: $N=50$; $\times$: $N=100$; $\bigcirc$:
$N=200$. Top pane is conditional on the true value of $P_{X}$ and
bottom pane is with $P_{X}$ integrated out.}
\label{Flo:var_vs_t}
\par\end{center}%
\end{minipage}\hfill{}%
\begin{minipage}[t]{0.49\columnwidth}%
\begin{center}
\includegraphics[width=1\textwidth]{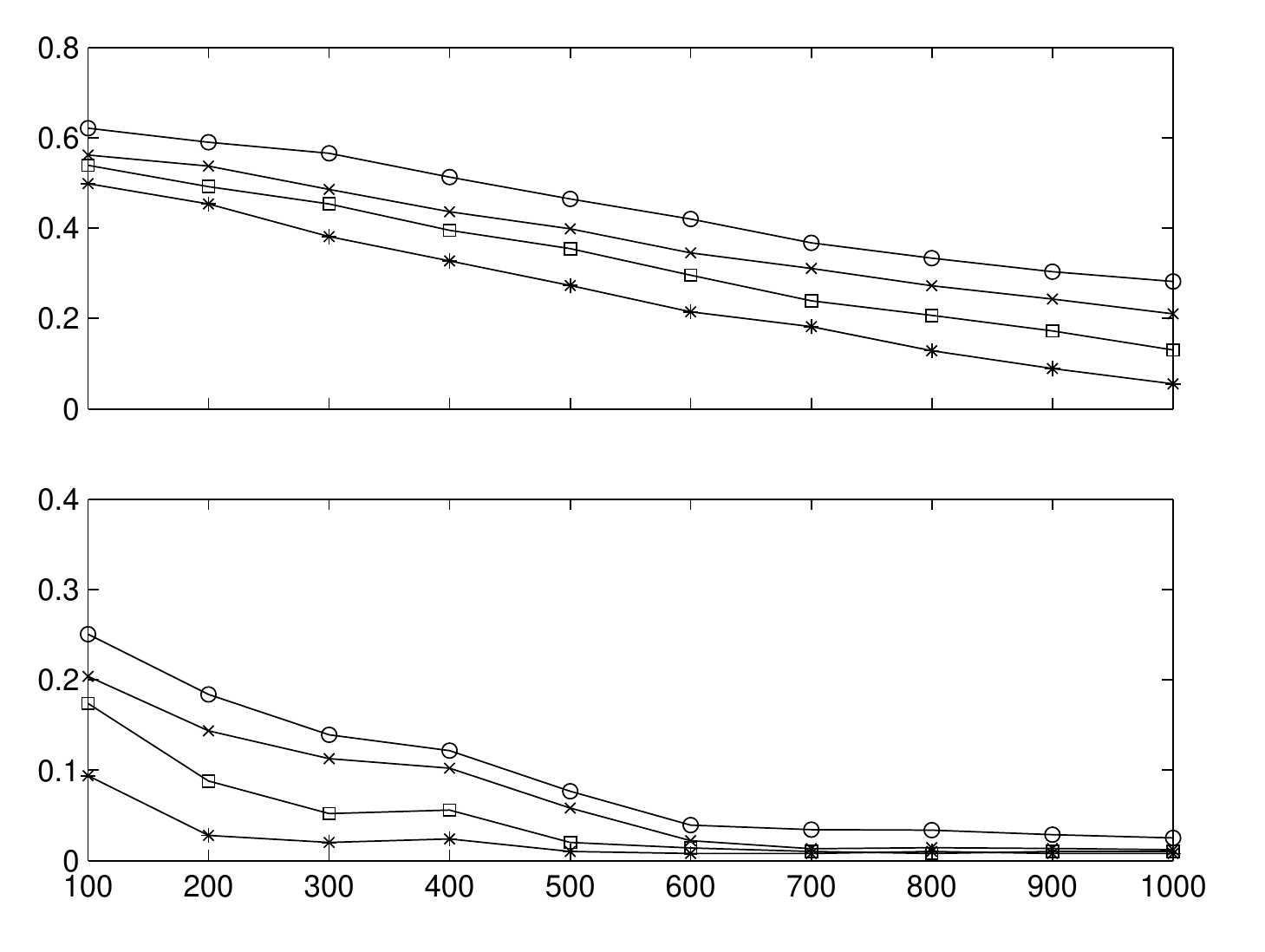}\caption{Example 1. PMMH acceptance rate as function of data record length
for various numbers of particles: blue: $N=10$, green: $N=50$, red:
$N=100$, black: $N=200$. Top pane is PMMH making proposals for $P_{X}$
and bottom pane is with $P_{X}$ integrated out.}
\label{Flo:a_rate_vs_t}
\par\end{center}%
\end{minipage}
\end{figure}

We found more obvious effects in the context of the PMMH algorithm,
which we now go on to discuss. In this case $f_{\theta}(x_{n}|x_{1:n-1})$
is computed as part of the DPF algorithm, which is where the same
conditioning issues arise. A data record of length $T=1000$ was generated
from the model with the same true parameter values as stated in the
previous section. The same prior distributions were also employed.
Central to the performance of the PMMH algorithm is the normalizing
constant estimate $\widehat{p}_{\theta}(y_{1:T})$ computed using
the DPF. When the variance of this estimate is large the PMMH algorithm
performs poorly, exhibiting a high rejection rate - a characteristic
shared with the standard PMCMC algorithms in \citet{andrieudoucetholenstein2008}.
We found that in the two cases (where $P_{X}$ was integrated out
and where it was not), the DPF exhibited striking differences in the
variance of this estimate. The parameter $\theta$ was set to its
true value and the DPF was run $1000$ times on the simulated data
set. Figure \ref{Flo:var_vs_t} shows the sample variance of $\log\widehat{p}_{\theta}(y_{1:n})$
as a function of $n$. The bottom pane corresponds to the case in
which $P_{X}$ is integrated out analytically. In this case the sample
variance grows super-linearly with $n$. By contrast, as shown in
the top pane, when conditioning on $P_{X}$ the variance grows far
more slowly. Very similar results were obtained when conditioning
on values of $P_{X}$ other than the truth. A step towards explaining
this phenomenon is noting that integrating out $P_{X}$ destroys the
ergodicity properties of the latent process $\left\{ X_{n}\right\} $
conditional on $\theta$. For standard SMC algorithms it is now theoretically
well understood that assumptions about the ergodicity properties of
the latent process are central to establishing linear growth rates
(with respect to $n$) for the error in normalizing constant-type
estimates \citep{Cer08}. Our numerical results are consistent with
the DPF having similar properties. 

\begin{figure}
\begin{centering}
\includegraphics[width=0.49\textwidth]{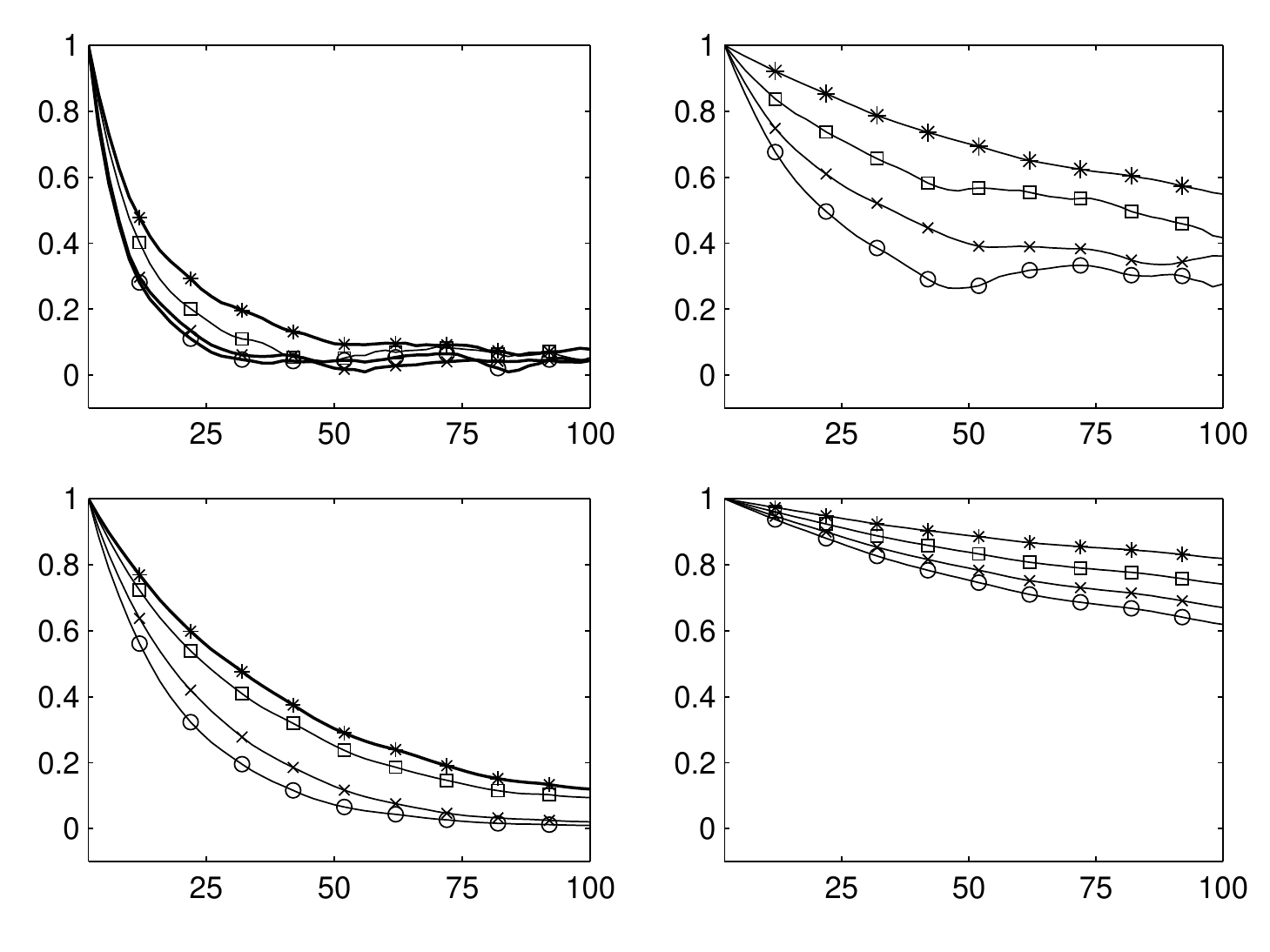}
\par\end{centering}

\caption{Example 1. Sample autocorrelation against lag for various numbers
of particles, $*$: $N=10$; $\square$: $N=50$, $\times$: $N=100$;
$\bigcirc$: $N=200$. Left column is PMMH with $P_{X}$ sampled and
right column is with $P_{X}$ integrated out analytically. Top plots
are for $\phi$ and bottom plots are for $\sigma^{2}$.}
\label{Flo:pmmh_ac}
\end{figure}

The variance of $\widehat{p}_{\theta}(y_{1:T})$ influences the acceptance
rates of the corresponding two PMMH algorithms. Of course the trade-off
is that when implementing a PMMH algorithm which incorporates $P_{X}$
into the sampling problem one has the added burden of designing proposal
moves for $P_{X}$ and the contribution to the variability of the
MH acceptance ratio from these proposals also influences the acceptance
rate. In our experiments we found that an effective approach to making
proposals for $P_{X}$ was to reparameterize the model in terms of
the unnormalized components of each row of $P_{X}$, with the Dirichlet
prior corresponding to gamma priors over these components. Proposals
could then be made using log-Gaussian random walks (an analogous approach
was advocated in the \citet{Jasra2005} in the context of static mixture
models). In numerical experiments we adopted this approach with independent
log-Gaussian random walk proposals made on each unnormalized component
of $P_{X}$. After a couple of preliminary runs, the standard deviation
of the increment in the log domain was set to $0.05$. A log-Gaussian
random walk proposal with the same standard deviation was also used
for the parameter $\sigma^{2}$ and a Gaussian random walk with standard
deviation $0.1$ was used for $\phi$. For the case where $P_{X}$
is integrated out we used the same proposals as above for $\phi$
and $\sigma^{2}$. Figure \ref{Flo:a_rate_vs_t} shows the PMMH acceptance
rates as a function of the length of the data record. These results
were obtained over $10^{5}$ iterations of the algorithms after a
burn-in of $10^{4}$. The results show that the acceptance rate drops
much more rapidly in the case that $P_{X}$ is integrated out. However,
we cannot conclude that the PMMH algorithm is always more efficient
when $P_{X}$ is incorporated into the sampling problem as the overall
efficiency naturally depends on the particular choice of proposal
mechanism for $P_{X}$. Our numerical results do indicate that even
using a fairly simple proposal mechanism for $P_{X}$ one can obtain
acceptance rates which are superior to those in the case that $P_{X}$
is integrated out analytically and the autocorrelation plots in Figure
\ref{Flo:pmmh_ac} show that this is carried over to lower sample
autocorrelation for the parameters $\phi$ and $\sigma^{2}$. 

\begin{figure}
\centering{}\includegraphics[width=0.5\columnwidth]{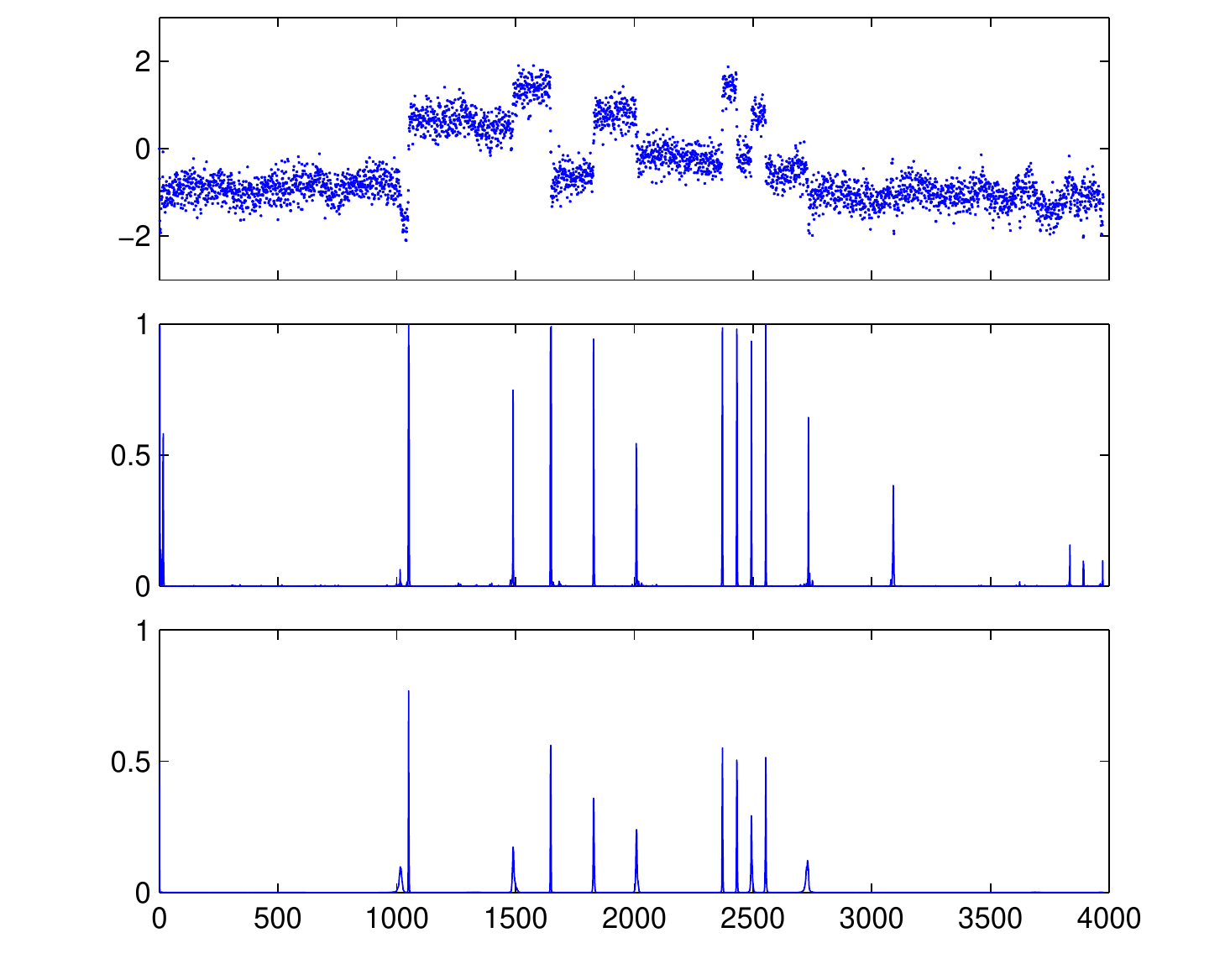}\caption{Example 2. Top: well-log data. Middle and bottom panes are estimated
$p(X_{n}=2|y_{1:T})$ from respectively the standard Gibbs and PG
samplers.}
\label{Flo:well_log_post}
\end{figure}

\subsection{Example 2: Multiple change-point model with dependence between segments.}

\label{subsec:welllogmodel} There is an extensive literature on statistical
time series analysis based on multiple change-point models. In such
models it is often assumed that given the position of a change-point,
the data after that change-point are conditionally independent of
those before, see for example \citet{BarryHart93,Fearn2007}, amongst
many others. This modelling assumption may be restrictive in some
circumstances. A natural way to relax it is via a SSSM, which allows
the notion of change-points to be introduced whilst allowing potentially
complex dependence structures across segments of the data.

We consider a multiple-change point model in which observations arise
from a latent process which is piece-wise linear. Changes in the latent
process are of two varieties: those in which there is a discontinuity
in the latent trajectory and its gradient and those in which there
is a discontinuity only in the gradient. More specifically, we have
$\mathcal{X}=\{0,1,2\}$ and we assume that $\left\{ X_{n}\right\} $
is Markov with unknown transition matrix $P_{X}.$ The observations
$\left\{ Y_{n}\right\} $ are real-valued, as are the latent trajectory
$\left\{ \mu_{n}\right\} $ and its gradient $\left\{ \dot{\mu}_{n}\right\} $.
In state-space form we have \[
Z_{n}=\left[\begin{array}{c}
\mu_{n}\\
\dot{\mu}_{n}\end{array}\right],\quad A_{\theta}(0)=\left[\begin{array}{cc}
1 & \Delta\\
0 & 1\end{array}\right],\quad A_{\theta}(1)=\left[\begin{array}{cc}
1 & \Delta\\
0 & 0\end{array}\right],\quad A_{\theta}(2)=\left[\begin{array}{cc}
0 & 0\\
0 & 0\end{array}\right],\]
\[
B_{\theta}(0)=\left[\begin{array}{cc}
0 & 0\\
0 & 0\end{array}\right],\quad B_{\theta}(1)=\left[\begin{array}{cc}
0 & 0\\
0 & \sigma_{\mu,1}\end{array}\right],\quad B_{\theta}(2)=\left[\begin{array}{cc}
\sigma_{\mu,0} & 0\\
0 & \sigma_{\mu,1}\end{array}\right],\]
\[
C_{\theta}(x_{n})=\left[\begin{array}{cc}
1 & 0\end{array}\right],\quad D_{\theta}(x_{n})=\sigma_{Y},\quad F_{\theta}(x_{n})=G_{\theta}(x_{n})=0,\quad\forall x_{n}.\]
Here $\Delta$ is a fixed time incremement and the unknown parameters
are $\theta=[\sigma_{Y}^{2}\;\sigma_{\mu,0}^{2}\;\sigma_{\mu,1}^{2}\; P_{X}]$.
We apply this model to the analysis of well-log data: measurements
of the nuclear resonance of underground rocks, as studied originally
in \citet{Ruanaidh96}. Observations arise from a drill bit which
passes down through layers of rock over time and each datum is a measurement
of the resonance of the rock through which the bit is passing at that
time. The aim is to identify segments in the data, each corresponding
to a stratum of a single type of rock. The data set we analyse was
treated in \citet{fearnhead2003,Fearn2007,Fearnhead2009} under a
variety of models, but in all these cases the static parameters of
the models were assumed known. In \citet{Fearnhead2009} a change-point
model with dependence across segments was employed and its advantages
in terms of avoiding spurious detection of change-points was demonstrated.
We are interested in similar analysis, but without assuming fixed
values for the static parameters of the model. As in \citet{fearnhead2003,Fearn2007,Fearnhead2009}
a few extreme outliers were removed from the data set manually resulting
in $3975$ data points. 

\begin{figure}
\begin{centering}
\includegraphics[width=0.7\columnwidth]{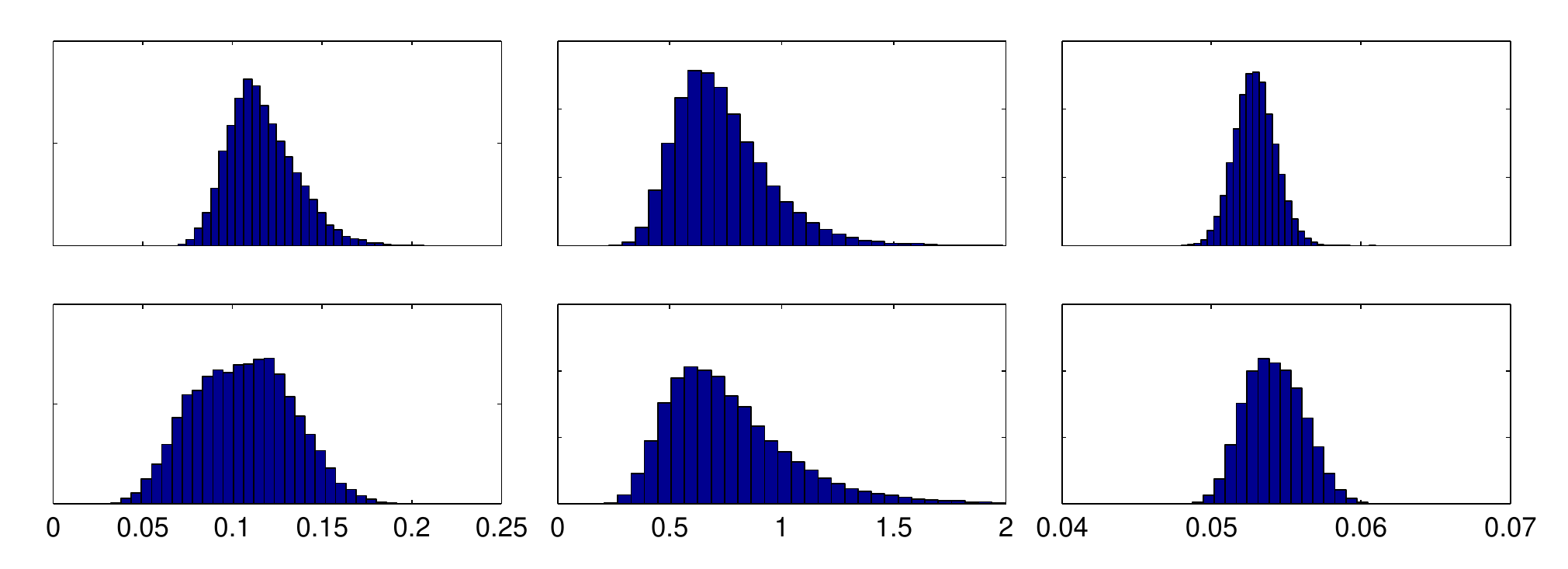}\caption{Example 2. Histograms estimates of posterior marginals. Top row: standard
Gibbs sampler. Bottom row: PG sampler. Columns from left to right
are $\sigma_{\mu,1}^{2}$, $\sigma_{\mu,0}^{2}$ and $\sigma_{Y}^{2}$. }
\label{Flo:well_log_hists}
\par\end{centering}

\end{figure}

Flat Dirichlet priors were set on each row of $P_{X}$. Independent
inverse gamma $(2,3)$ priors were placed over $\sigma_{Y}^{2}$,
$\sigma_{\mu,0}^{2}$ and $\sigma_{\mu,1}^{2}$. In our experiments,
inference was found to be insensitive to choice of parameters for
these inverse gamma priors (not shown). For the initial distribution
over $Z_{0}$ we set a relatively diffuse, zero mean Gaussian prior
with diagonal covariance components $100$ and $100$, corresponding
to $\mu_{0}$ and $\dot{\mu}_{0}$ respectively. We set $\Delta=0.1$.
The standard Gibbs sampler was run for $2\times10^{6}$ iterations
and PG sampler with $N=50$ for $4\times10^{4}$ iterations so as
to equate computational cost. Histograms of sample output for $\sigma_{Y}^{2}$,
$\sigma_{\mu,0}^{2}$ and $\sigma_{\mu,1}^{2}$ are shown in Figure
\ref{Flo:well_log_hists}. These results indicate that despite the
long run the standard Gibbs sampler has not converged: most noticeably
in the case of the histograms for $\sigma_{\mu,1}^{2}$, it appears
not to have explored the support as thoroughly as the PG sampler and
has become stuck in a mode of the distribution. The difference in
performance is even more striking when considering the corresponding
estimated posterior probabilities for the latent switching process.
Figure \ref{Flo:well_log_post} shows the estimated marginal posterior
probabilities of each $X_{n}$ being in state $2$ (recall this state
corresponds to a discontinuity in the latent process $\left\{ \mu_{n}\right\} $
and its gradient) for each time step of the data record. Due to the
lack of full exploration of the parameter space, the results for the
standard Gibbs sampler show erroneously high posterior probabilities
that each $X_{n}$ is in state 2. We can conclude that for the same
computational cost the performance of the PG sampler is superior.

\subsection{Example 3: Exchange Rate Model}

\label{subsec:ex_rate_model} The following model was investigated
in \citet{engle1999,fruwirthschnatter2006}, where it was used to
analyze economic data. The model consists of a latent random walk
component observed in auto-regressive noise, where the variance of
the observation noise innovations can switch between different values.
In \citet{engle1999}, this model was advocated to reflect the heteroscedasticity
evident in the price index adjusted U.S./U.K. exchange rate during
the late 19th and 20th centuries. The data consist of $1322$ monthly
log exchange rate values. We consider the case treated in \citet{fruwirthschnatter2006}
where the auto-regressive noise process is of order $2$ and there
are $4$ switching states. In this model, $\mathcal{X}=\{1,2,3,4\}$
and the discrete latent process $\{X_{n}\}$ is a Markov chain with
transition matrix $P_{X}$. The observations $\{Y_{n}\}$ are log-exchange
rate values. The latent process $\left\{ \mu_{n}\right\} $ is a random
walk and we denote by $\left\{ \eta_{n}\right\} $ the auto-regressive
noise process: \begin{align*}
Y_{n} & =\mu_{n}+\eta_{n},\\
\mu_{n} & =\mu_{n-1}+\sigma_{\mu}V_{n,1}\\
\eta_{n} & =a_{1}\eta_{n-1}+a_{2}\eta_{n-2}+\sigma_{\eta,X_{n}}V_{n,2}\end{align*}
 where $\{V_{n,1}\}$ and $\{V_{n,2}\}$ are i.i.d. $\mathcal{N}(0,1)$
noise sequences. In state-space form we then have\[
Z_{n}=\left[\begin{array}{c}
\mu_{n}\\
\eta_{n}\\
\eta_{n-1}\end{array}\right],\quad A_{\theta}(x_{n})=\left[\begin{array}{ccl}
1 & 0 & 0\\
0 & a_{1} & a_{2}\\
0 & 1 & 0\end{array}\right],\quad B_{\theta}(x_{n})=\left[\begin{array}{ccl}
\sigma_{\mu} & 0 & 0\\
0 & \sigma_{\eta,x_{n}} & 0\\
0 & 0 & 0\end{array}\right],\]
\[
C_{\theta}(x_{n})=\left[\begin{array}{ccl}
1 & 1 & 0\end{array}\right],\quad D_{\theta}(x_{n})=F_{\theta}(x_{n})=G_{\theta}(x_{n})=0,\quad\forall x_{n}.\]
The unknown parameters of the model are $\theta=[\sigma_{\mu,}^{2}\;\sigma_{\eta,1}^{2}\;\sigma_{\eta,2}^{2}\;\sigma_{\eta,3}^{2}\;\sigma_{\eta,4}^{2}\; a_{1}\; a_{2}\; P_{X}]$.
Under symmetric priors the labeling of the discrete states is not
identifiable. We consider the same prior distributions on the parameters
and initial conditions on $Z_{0}$ as in \citet{fruwirthschnatter2006}
and we refer to the latter for full details, including a stability
constraint on the auto-regressive coefficients $(a_{1,}a_{2})$. The
only difference is that we do not impose an identifiability constraint
a priori on $\sigma_{\eta,1}^{2},\sigma_{\eta,2}^{2},\sigma_{\eta,3}^{2},\sigma_{\eta,4}^{2}$,
but instead target the unidentified model and impose the ordering
$\sigma_{\eta,1}^{2}<\sigma_{\eta,2}^{2}<\sigma_{\eta,3}^{2}<\sigma_{\eta,4}^{2}$
after sampling (see \citet{fruwirthschnatter2001,Jasra2005} and references
therein for various approaches to drawing inference in models with
unidentifiable state labels). 

\begin{figure}
\includegraphics[width=0.5\textwidth]{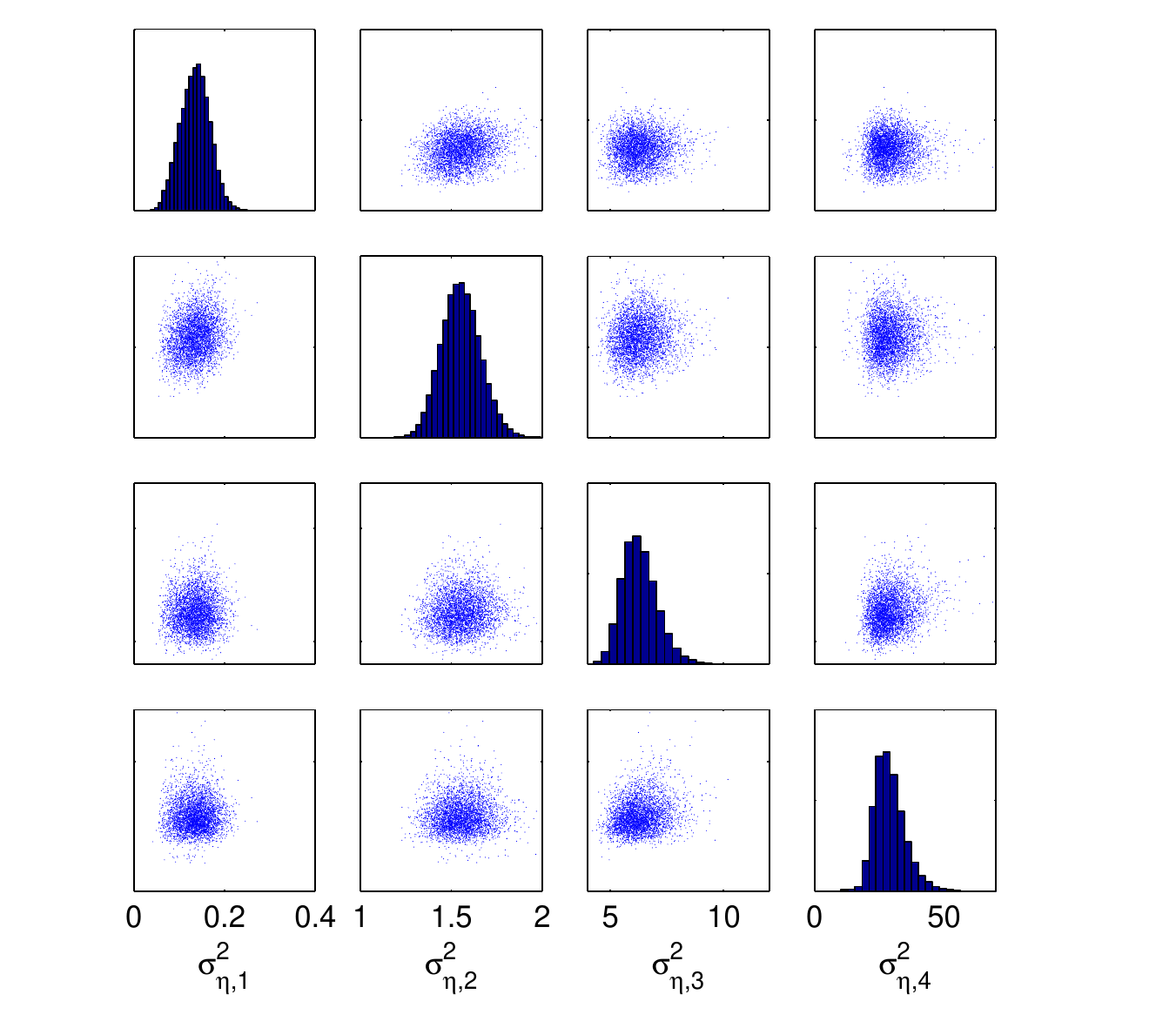}\hfill{}\includegraphics[width=0.5\textwidth]{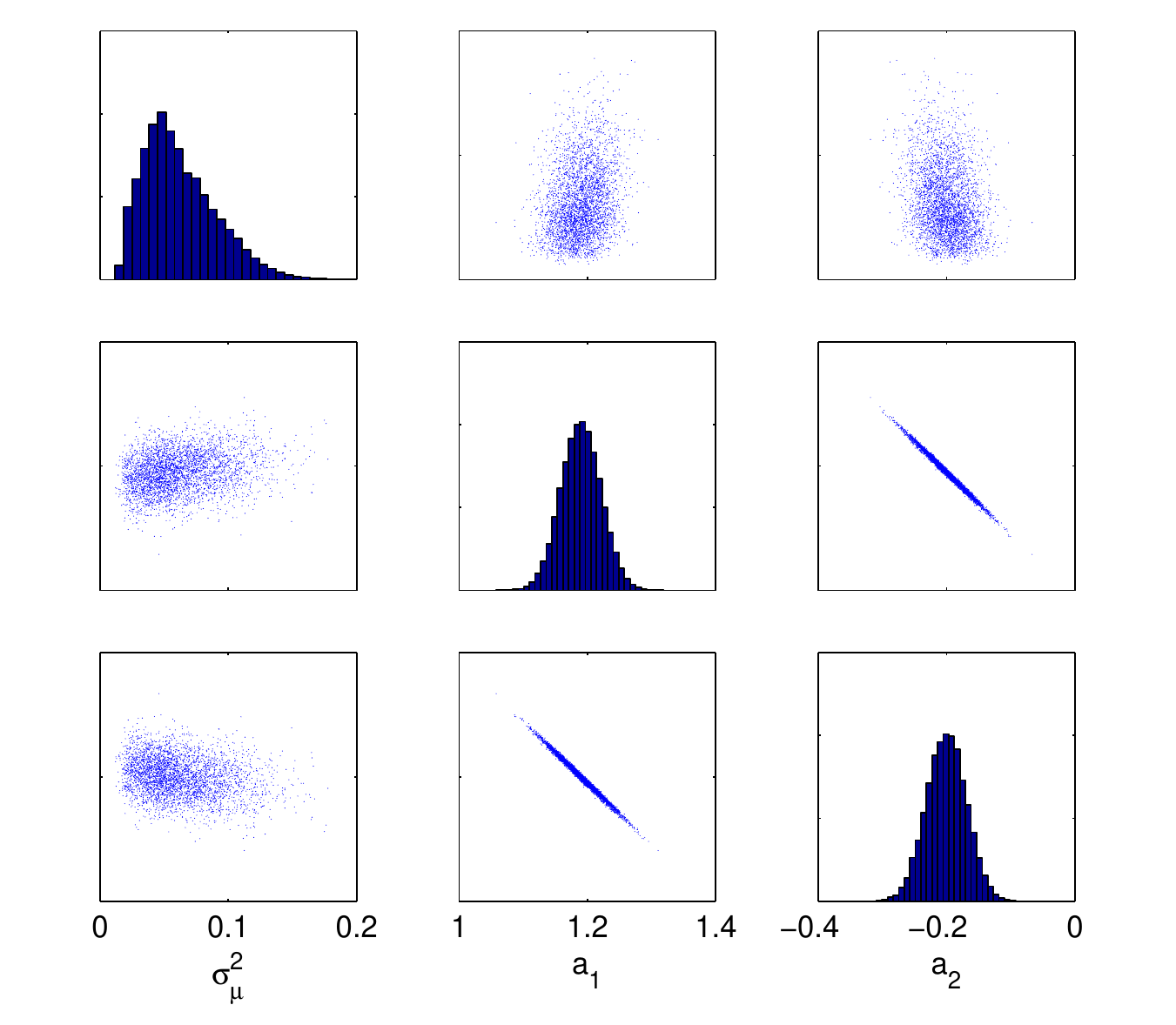}\caption{Example 3. Histogram estimates of posterior marginals and scatter
plots of pairwise marginals for the exchange rate model.}
\label{Flo:ex_rate_hists1}

\end{figure}

\begin{figure}
\begin{centering}
\includegraphics[width=0.5\textwidth]{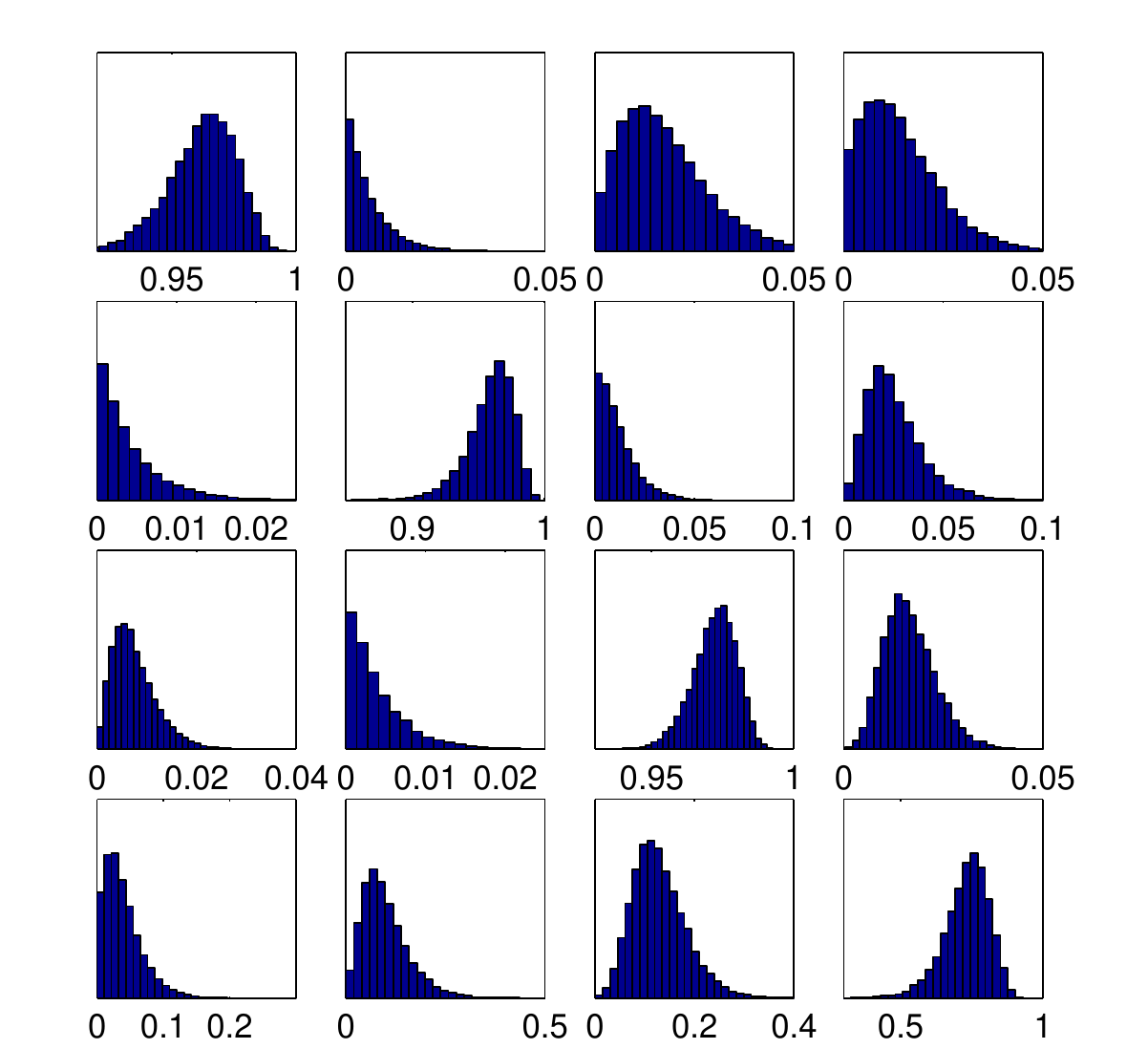}
\par\end{centering}

\caption{Example 3. Histogram estimates of marginal posterior distributions
for entries of the state transition matrix $P_{X}$. Panes are arranged
as per the transition matrix itself.}
\label{Flo:ex_rate_hists2}

\end{figure}

\begin{figure}
\begin{centering}
\includegraphics[width=0.9\textwidth]{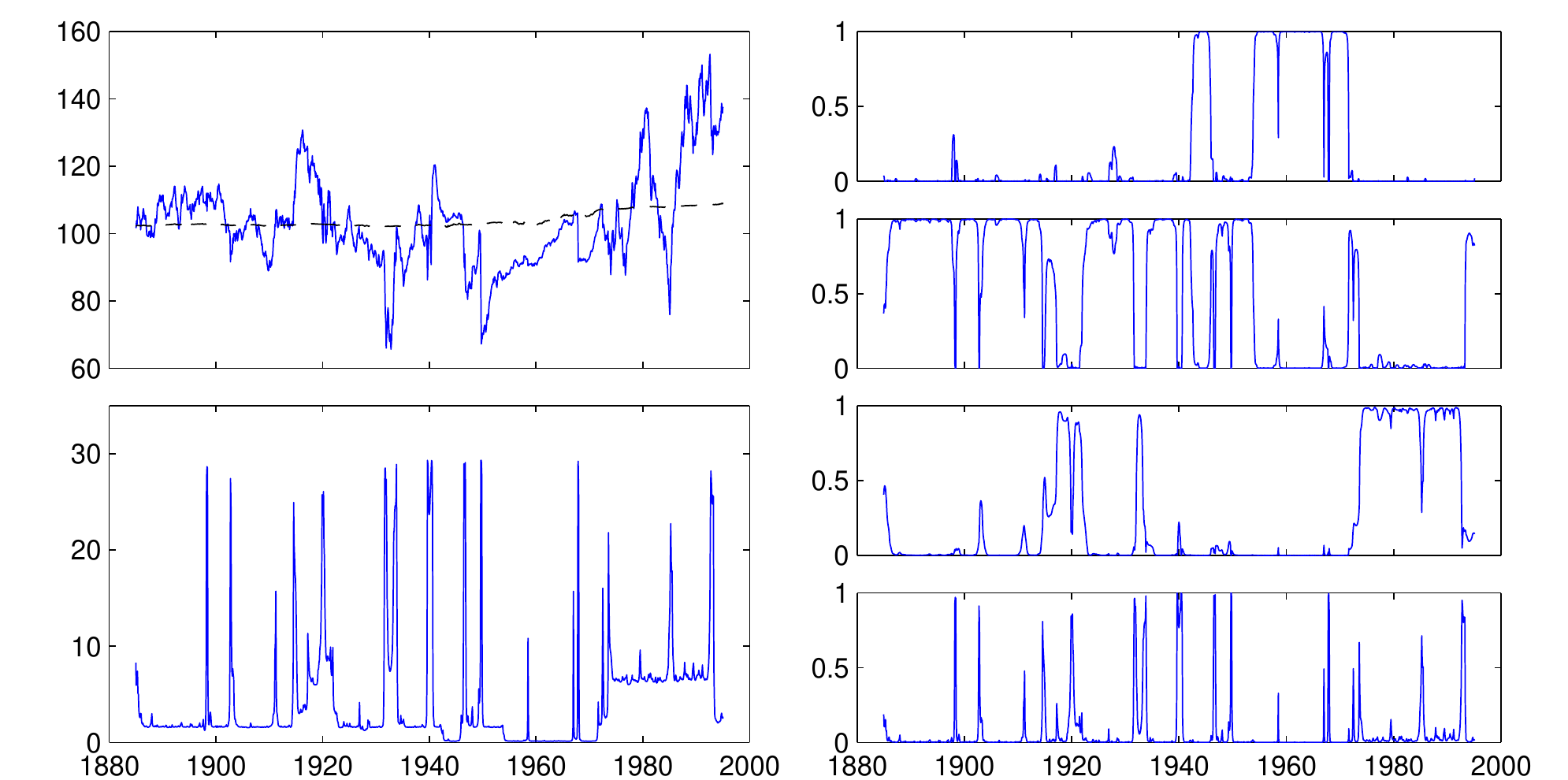}
\par\end{centering}

\caption{Example 3. Top left: data (solid) and $\mathbb{E}\left[\left.\mu_{n}\right|y_{1:T}\right]$
(dashed). Bottom left: $\mathbb{E}\left[\left.\sigma_{\eta,X_{n}}^{2}\right|y_{1:T}\right]$.
Right: estimated posterior probabilities $p(X_{n}=j|y_{1:T})$ for,
top to bottom, $j=1,2,3,4$. }
\label{Flo:ex_rate_post}
\end{figure}

We implemented an algorithm for this model with $P_{X}$ incorporated
into the sampling. Each iteration of the algorithm consisted of a
sequence of two PMMH updates. The first holding $P_{X}$ and $\sigma_{\eta,1}^{2},\sigma_{\eta,2}^{2},\sigma_{\eta,3}^{2},\sigma_{\eta,4}^{2}$
constant and the second holding $(a_{1,}a_{2})$ and $\sigma_{\mu}^{2}$
constant (using standard arguments for Metropolis-within-Gibbs algorithms
and Theorem \ref{theorem:particleMMH} it is straightforward to show
this sequence of updates is invariant with respect to the extended
target distribution). After a couple of preliminary runs the following
proposals were selected. A symmetric random walk proposal of standard
deviation of $0.001$ was used for $(a_{1,}a_{2})$ and for $\sigma_{\mu}^{2}$
a log-Gaussian random walk with log-domain standard deviation of $0.01$.
We used a mixture of log-Gaussian random walks for the unnormalised
components of $P_{X}$ and $\sigma_{\eta,1}^{2},\sigma_{\eta,2}^{2},\sigma_{\eta,3}^{2},\sigma_{\eta,4}^{2}$.
For each individual parameter, the mixture had two components, the
first with weight $0.9$ and standard deviation $0.05$ in the log
domain and the second with weight $0.1$ and standard deviation $1$
in the log domain. With these settings and $N=200$ we achieved an
overall acceptance rate of $0.2$. This is a reasonable rate given
the mixture proposals. The algorithm was run for $2\times10^{5}$
iterations after an initial burn-in of $10^{4}$. Inferential summaries
are presented in Figures \ref{Flo:ex_rate_hists1}-\ref{Flo:ex_rate_post}.
We note that there are some differences between the results we obtained
and those from \citet{fruwirthschnatter2006}, where a standard Gibbs
sampler was applied. We conjecture that the latter had not fully explored
the support of the posterior distribution. Noticeable differences
are that the posterior marginal for $\sigma_{\eta,1}^{2}$ we obtain
is more diffuse than that reported in \citet{fruwirthschnatter2006}
and we obtain a much flatter trajectory in the posterior estimates
of $\{\mu_{n}\}$ in Figure \ref{Flo:ex_rate_post}. Another significant
difference is that we obtain concentration of the marginal posterior
over the auto-regressive coefficients $(a_{1},a_{2})$ in a different
region than that reported in \citet{fruwirthschnatter2006}. Using
other proposals for $(a_{1},a_{2})$ we were not able to find another
major mode. Furthermore the posterior marginal for $\sigma_{\mu}^{2}$
we obtained is concentrated on lower values. Overall, we feel that
the ability to integrate out approximately the latent variables makes
the PMMH algorithm a powerful tool: as these results demonstrate it
gives us the chance to explore regions of posterior support which
Gibbs sampling algorithms may struggle to find.

\section{Discussion and extensions\label{sec:discussion_extensions}}

In this article, we have proposed new PMCMC algorithms relying on
the efficient DPF algorithm to perform Bayesian inference in SSSM.
We have shown experimentally that these generic discrete PMCMC algorithms
outperform current state-of-the-art MCMC techniques for a given computational
complexity. Moreover the DPF\ can be easily parallelised so further
substantial improvements could be obtained.

There are various possible extensions to this work. First, we have
restricted ourselves to SSSM but the DPF can be applied to any model
where the latent process is discrete-valued. This includes for example
Dirichlet process mixtures \citep{Fearnhead2004} and the infinite
hidden Markov model introduced in \citet{teh2006}. Compared to the
SSSM framework, the differences are that, in these scenarios, $X_{n}$
takes values in a set whose cardinality increases over time and computations
required to evaluate the importance weights are not performed using
the Kalman filter. However, the discrete PMCMC methodology discussed
here can be straightforwardly extended to these cases. Second, it
would be possible to extend the DPF and the associated discrete PMCMC
methodology by using look-ahead techniques. In a look-ahead strategy
with an integer lag $L$, we resample trajectories at time $n$ by
considering the weights proportional to $p_{\theta}\left(x_{1:n}|y_{1:n+L}\right)$
instead of $p_{\theta}\left(x_{1:n}|y_{1:n}\right)$ for the standard
DPF. This is obviously more expensive than the DPF as computation
of the weights involves summing over $x_{n+1:L}$ for each particle,
but this might be of interest in scenarios where future observations
are very informative about $X_{n}$.

\newpage{}

\appendix

\section{Kalman Filter\label{sec:kalman}}

Conditional upon $X_{1:T}=x_{1:T}$, Eq. (\ref{eq:evollineargauss})-(\ref{eq:obslineargauss})
defines a linear Gaussian state-space model. The Kalman filter allows
us to compute recursively in time $p_{\theta}\left(\left.z_{n}\right\vert y_{1:n-1},x_{1:n}\right)=\mathcal{N}\left(z_{n};m_{\left.n\right\vert n-1}^{z,\theta}\left(x_{1:n}\right),\Sigma_{\left.n\right\vert n-1}^{z,\theta}\left(x_{1:n}\right)\right)$,
$p_{\theta}\left(\left.z_{n}\right\vert y_{1:n},x_{1:n}\right)=\mathcal{N}\left(z_{n};m_{\left.n\right\vert n}^{z,\theta}\left(x_{1:n}\right),\Sigma_{\left.n\right\vert n}^{z,\theta}\left(x_{1:n}\right)\right)$
and the predictive density\\
$g_{\theta}\left(\left.y_{n}\right\vert y_{1:n-1},x_{1:n}\right)=\mathcal{N}\left(y_{n};m_{\left.n\right\vert n-1}^{y,\theta}\left(x_{1:n}\right),\Sigma_{\left.n\right\vert n-1}^{y,\theta}\left(x_{1:n}\right)\right)$.
For $n\geq1$ these statistics are computed using the following recursion
initialized with $m_{\left.0\right\vert 0}^{z}=m_{0},$ $\Sigma_{\left.0\right\vert 0}^{z}=\Sigma_{0}$
\begin{align*}
m_{\left.n\right\vert n-1}^{z,\theta}\left(x_{1:n}\right) & =A_{\theta}(x_{n})m_{\left.n-1\right\vert n-1}^{z}\left(x_{1:n-1}\right)+F_{\theta}(x_{n})u_{n},\text{ }\\
\Sigma_{\left.n\right\vert n-1}^{z,\theta}\left(x_{1:n}\right) & =A_{\theta}(x_{n})\Sigma_{\left.n-1\right\vert n-1}^{z}A_{\theta}^{\text{T}}(x_{n})+B_{\theta}(x_{n})B_{\theta}^{\text{T}}(x_{n}),\\
m_{\left.n\right\vert n-1}^{y,\theta}\left(x_{1:n}\right) & =C_{\theta}(x_{n})m_{\left.n\right\vert n-1}^{z,\theta}\left(x_{1:n}\right)+G_{\theta}(x_{n})u_{n},\\
\Sigma_{\left.n\right\vert n-1}^{y,\theta}\left(x_{1:n}\right) & =C_{\theta}(x_{n})\Sigma_{\left.n\right\vert n-1}^{z,\theta}\left(x_{1:n}\right)C_{\theta}^{\text{T}}(x_{n})+D_{\theta}(x_{n})D_{\theta}^{\text{T}}(x_{n}),\\
m_{\left.n\right\vert n}^{z,\theta}\left(x_{1:n}\right) & =m_{\left.n\right\vert n-1}^{z,\theta}\left(x_{1:n}\right)+\Sigma_{\left.n\right\vert n-1}^{z,\theta}\left(x_{1:n}\right)C_{\theta}^{\text{T}}(x_{n})\left[\Sigma_{\left.n\right\vert n-1}^{y,\theta}\left(x_{1:n}\right)\right]^{-1}\left(y_{n}-m_{\left.n\right\vert n-1}^{y,\theta}\left(x_{1:n}\right)\right),\\
\Sigma_{\left.n\right\vert n}^{z,\theta}\left(x_{1:n}\right) & =\Sigma_{\left.n\right\vert n-1}^{z,\theta}\left(x_{1:n}\right)-\Sigma_{\left.n\right\vert n-1}^{z,\theta}\left(x_{1:n}\right)C_{\theta}^{\text{T}}(x_{n})\left[\Sigma_{\left.n\right\vert n-1}^{y,\theta}\left(x_{1:n}\right)\right]^{-1}C_{\theta}\left(x_{n}\right)\Sigma_{\left.n\right\vert n-1}^{z,\theta}\left(x_{1:n}\right).\end{align*}

\section{Backward Sampling\label{sec:Backward-Sampling}}

A key component of the backward sampling algorithm is the evaluation
of the backward weight 

\[
V_{n}^{\theta}\left(x_{1:n}\left|x_{n+1:T}'\right.\right)\propto W_{n}^{\theta}(x_{1:n})p_{\theta}(x_{n+1:T}'|x_{1:n})p_{\theta}(y_{n+1:T}|y_{1:n},x_{1:n},x_{n+1:T}')\]
for each candidate sub-trajectory $x_{1:n}$ and where $x_{n+1:T}'$
is the complementing sub-trajectory which has been obtained from previous
steps of the backward sampling procedure. Central to the computation
of this weight is the identity

\begin{equation}
p_{\theta}(y_{n+1:T}|y_{1:n},x_{1:n},x_{n+1:T}')=\int p_{\theta}(y_{n+1:T}|z_{n},x_{n+1:T}')p_{\theta}(z_{n}|x_{1:n},y_{1:n})dz_{n},\label{eq:backward_id_appendix}\end{equation}
where $p_{\theta}(z_{n}|x_{1:n},y_{1:n})$ is the Gaussian conditional
filtering density associated with the sub-trajectory $x_{1:n}$ and
is specified by its mean vector $m_{\left.n\right\vert n}^{z,\theta}\left(x_{1:n}\right)$
and co-variance matrix $\Sigma_{\left.n\right\vert n}^{z,\theta}\left(x_{1:n}\right)$.
In order to compute (\ref{eq:backward_id_appendix}) (at least up
to a constant of proportionality) it is necessary to obtain the coefficients
of $z_{n}$ in $p_{\theta}(y_{n+1:T}|z_{n},x_{n+1:T}')$. The latter
can be expressed as 

\[
p_{\theta}(y_{n+1:T}|z_{n},x_{n+1:T}')\propto\exp\left[-\frac{1}{2}\left(z_{n}^{T}\Xi_{n}z_{n}-2\mu_{n}^{T}z_{n}\right)\right]\]
where $\Xi_{n}$ and $\mu_{n}$ are respectively a matrix and vector
of appropriate dimension, both depending on $x_{n+1:T}'$, $y_{n+1:T}$
and $\theta$. In the following this dependence is suppressed from
the notation for convenience. For ease of presentation we use the
similarly abusive conventions in writing $m_{n}=m_{\left.n\right\vert n}^{z,\theta}\left(x_{1:n}\right)$,
$\Sigma_{n}=\Sigma_{\left.n\right\vert n}^{z,\theta}\left(x_{1:n}\right)$,
$A_{n}=A_{\theta}(x_{n})$, $B_{n}=\left[B_{\theta}(x_{n})\;\;0_{z\times w}\right]$,
$C_{n}=C_{\theta}(x_{n})$, $D_{n}=\left[0_{y\times v}\;\; D_{\theta}(x_{n})\right],$
$F_{n}=F_{\theta}(x_{n})$, $G_{n}=G_{\theta}(x_{n})$. Then let $\Upsilon_{n}$
be a matrix satisfying $\Sigma_{\left.n\right\vert n}^{z,\theta}\left(x_{1:n}\right)=\Upsilon_{n}\Upsilon_{n}^{T}$.
We have

\begin{eqnarray}
 &  & p_{\theta}(y_{n+1:T}|y_{1:n},x_{1:n},x_{n+1:T}')\nonumber \\
 &  & \propto\exp\left(-\frac{1}{2}\left[m_{n}^{T}\Xi_{n}m_{n}-2\mu_{n}^{T}m_{n}-\left(\mu_{n}-\Xi_{n}m_{n}\right)^{T}\Upsilon_{n}\left(\Upsilon_{n}\Xi_{n}\Upsilon_{n}+I\right)^{-1}\Upsilon_{n}^{T}\left(\mu_{n}-\Xi_{n}m_{n}\right)\right]\right)\nonumber \\
 &  & \quad\times\left|\Upsilon_{n}^{T}\Xi_{n}\Upsilon_{n}+I\right|^{-1/2}.\label{eq:backward_weight_like}\end{eqnarray}
where $I$ is the identity matrix of appropriate dimension. We now
specify equations for updating $(\mu_{n},\Xi_{n})$, which are given
without proof of validity: they are a direct application of Lemmata
1 and 2 in \citet{gerlach2000}. As in \citet{gerlach2000}, for simplicity
we present recursions only for the case in which the observations
are scalar-valued, but they can readily be extended to the vector-valued
case. Let \begin{align*}
r_{n+1}= & \left(C_{n+1}B_{n+1}+D_{n+1}\right)\left(C_{n+1}B_{n+1}+D_{n+1}\right)^{T},\\
\Phi_{n+1}= & B_{n+1}\left(B_{n+1}^{T}C_{n+1}^{T}+D_{n+1}^{T}\right)/r_{n+1},\\
\Lambda_{n+1}= & \left(1-\Phi_{n+1}C_{n+1}^{T}\right)A_{n+1},\\
a_{n+1}= & \left(1-\Phi_{n+1}C_{n+1}^{T}\right)F_{n+1}u_{n+1}-\Phi_{n+1}G_{n+1}u_{n+1},\end{align*}
 and let $\Gamma_{n+1}$ be a matrix which satisfies \[
\Gamma_{n+1}\Gamma_{n+1}^{T}=B_{n+1}\left(I-\frac{1}{r_{n+1}}\left(B_{n+1}^{T}C_{n+1}^{T}+D_{n+1}^{T}\right)\left(B_{n+1}^{T}C_{n+1}^{T}+D_{n+1}^{T}\right)^{T}\right)B_{n+1}^{T}.\]
 The recursion for $(\mu_{n},\Xi_{n})$ is then given by
\begin{itemize}
\item Set $\Xi_{T}=0$, $\mu_{T}=0$.
\item For $n=T-1,...,1$\begin{eqnarray*}
M_{n+1} & = & \Gamma_{n+1}^{T}\Xi_{n+1}\Gamma_{n+1}+I,\\
\\\Xi_{n} & = & \Lambda_{n+1}^{T}\left(\Xi_{n+1}-\Xi_{n+1}\Gamma_{n+1}M_{n+1}^{-1}\Gamma_{n+1}^{T}\Xi_{n+1}\right)\Lambda_{n+1}+A_{n+1}^{T}C_{n+1}^{T}C_{n+1}A_{n+1}\frac{1}{r_{n+1}},\\
\\\mu_{n} & = & \Lambda_{n+1}^{T}\left(I-\Xi_{n+1}\Gamma_{n+1}M_{n+1}^{-1}\Gamma_{n+1}^{T}\right)\left(\mu_{n+1}-\Xi_{n+1}\left(a_{n+1}+\Phi_{n+1}y_{n+1}\right)\right)\\
 &  & +A_{n+1}^{T}C_{n+1}^{T}\left(y_{n+1}-G_{n+1}u_{n+1}-C_{n+1}F_{n+1}u_{n+1}\right)\frac{1}{r_{n+1}}.\end{eqnarray*}

\end{itemize}

\section{Proofs\label{app:Proofs}}

\noindent \textbf{Proof of Theorem} \ref{theorem:particleMMH}\textbf{.}
We obtain from Eq. (\ref{eq:propertymarginalresampling})-(\ref{eq:DPFsamplingdistribution})-(\ref{eq:artificialdistributiondetails})
that on the event $x_{1:T}\in\mathbf{S}_{T}$, \begin{align*}
\frac{\pi_{\theta}^{N}(x_{1:T},\mathbf{s}_{1},\mathbf{s}_{2},...,\mathbf{s}_{T})}{w_{T}^{\theta}\left(x_{1:T}\right)\text{ }\psi_{\theta}^{N}\left(\mathbf{s}_{1},\mathbf{s}_{2},...,\mathbf{s}_{T}\right)} & =\frac{p_{\theta}(\left.x_{1:T}\right\vert y_{1:T})\left\{ \prod_{n=2}^{T}\mathbb{I}[x_{1:n}\in\mathbf{s}_{n}]\right\} }{w_{T}^{\theta}\left(x_{1:T}\right)\prod_{n=2}^{T}r_{n}^{N}(x_{1:n}\in\mathbf{s}_{n}|\mathbf{w}_{n-1}^{\theta})}\\
 & =\frac{p_{\theta}(\left.x_{1:T}\right\vert y_{1:T})\left\{ \prod_{n=2}^{T}\mathbb{I}[x_{1:n}\in\mathbf{s}_{n}]\right\} }{w_{T}^{\theta}\left(x_{1:T}\right)\prod_{n=1}^{T-1}\left(1\wedge C_{n}w_{n}^{\theta}\left(x_{1:n}\right)\right)}.\end{align*}
 It follows from Eq. (\ref{eq:updateweight1})-(\ref{eq:updateweight2})
that on the event $x_{1:T}\in\mathbf{S}_{T}$ the normalized weight
can be expanded as follows \begin{align}
w_{T}^{\theta}\left(x_{1:T}\right) & =\nu_{\theta}(x_{1})g_{\theta}(y_{1}|x_{1})\prod_{n=2}^{T}f_{\theta}(x_{n}|x_{1:n-1})g_{\theta}(y_{n}|y_{1:n-1},x_{1:n})\nonumber \\
 & \times\prod_{n=1}^{T-1}\frac{1}{1\wedge C_{n}w_{n}^{\theta}\left(x_{1:n}\right)}\text{ }\prod_{n=1}^{T}\left\{ \sum_{x_{1:n}^{\prime}\in\mathbf{s}_{n}}\overline{w}_{n}^{\theta}\left(x_{1:n}^{\prime}\right)\right\} ^{-1}\label{eq:weightexpansion}\end{align}
 Hence, using Eq.\ (\ref{eq:marginallikelihoodSMC})-(\ref{eq:marginallikelihoodincrementsSMC}),
we obtain \begin{equation}
\frac{\pi_{\theta}^{N}(x_{1:T},\mathbf{s}_{1},\mathbf{s}_{2},...,\mathbf{s}_{T})}{w_{T}^{\theta}\left(x_{1:T}\right)\text{ }\psi_{\theta}^{N}\left(\mathbf{s}_{1},\mathbf{s}_{2},...,\mathbf{s}_{T}\right)}=\frac{\widehat{p}_{\theta}\left(y_{1:T}\right)}{p_{\theta}\left(y_{1:T}\right)}.\label{eq:resultratio}\end{equation}
From (\ref{eq:resultratio}), we can now easily establish that an
MH sampler of target density (\ref{eq:artificialdistributiondetails})
and proposal density (\ref{eq:proposalPMMH}) admits indeed Eq. (\ref{eq:acceptanceprobaPMMH})
as MH ratio and the first part of the theorem follows. The second
part of the proof is a direct consequence of Theorem 1 in \citet{andrieuroberts2006}
and (A\ref{hyp:trueMHconverges}). $\blacksquare$

\noindent \textbf{Proof of Proposition} \ref{prop:backward_sampling}.
First note that, from Eq. (\ref{eq:artificialdistributiondetails})
and Eq. (\ref{eq:weightexpansion}), for $\theta,\mathbf{s}_{1},\mathbf{s}_{2},...,\mathbf{s}_{T}$
in the support of $\pi^{N}(\theta,\mathbf{s}_{1},\mathbf{s}_{2},...,\mathbf{s}_{T})$,

\begin{eqnarray}
\pi^{N}(x_{1:T}|\theta,\mathbf{s}_{1},\mathbf{s}_{2},...,\mathbf{s}_{T}) & \propto & p_{\theta}(\left.x_{1:T}\right\vert y_{1:T})\left\{ \prod_{n=2}^{T}\mathbb{I}[x_{1:n}\in\mathbf{s}_{n}]\right\} \frac{\psi_{\theta}^{N}(\mathbf{s}_{1},\mathbf{s}_{2},...,\mathbf{s}_{T})}{\prod_{n=2}^{T}r_{n}^{N}(x_{1:n}\in\mathbf{s}_{n}|\mathbf{w}_{n-1}^{\theta})}\nonumber \\
 & \propto & \nu_{\theta}(x_{1})g_{\theta}(y_{1}|x_{1})\prod_{n=2}^{T}f_{\theta}(x_{n}|x_{1:n-1})g_{\theta}(y_{n}|y_{1:n-1},x_{1:n})\nonumber \\
 &  & \times\left\{ \prod_{n=2}^{T}\mathbb{I}[x_{1:n}\in\mathbf{s}_{n}]\right\} \prod_{n=1}^{T-1}\frac{1}{1\wedge c_{n}w_{n}^{\theta}\left(x_{1:n}\right)}\label{eq:backward_samp_prop_W}\\
 & \propto & w_{T}^{\theta}(x_{1:T}).\nonumber \end{eqnarray}
Furthermore, for $1\leq n\leq T-1$,

\begin{eqnarray}
\pi^{N}(x_{1:n}|\theta,x_{n+1:T},\mathbf{s}_{1},\mathbf{s}_{2},...,\mathbf{s}_{n}) & \propto & p_{\theta}(x_{1:T}|y_{1:T})\left\{ \prod_{k=2}^{n}\mathbb{I}[x_{1:k}\in\mathbf{s}_{k}]\right\} \frac{\psi_{\theta}^{N}(\mathbf{s}_{1},\mathbf{s}_{2},...,\mathbf{s}_{n})}{\prod_{k=2}^{n}r_{k}^{N}(x_{1:k}\in\mathbf{s}_{k}|\mathbf{w}_{k-1}^{\theta})}\nonumber \\
 & \propto & \frac{p_{\theta}(x_{1:n}|y_{1:n})}{\prod_{k=2}^{n}r_{k}^{N}(x_{1:k}\in\mathbf{s}_{k}|\mathbf{w}_{k-1}^{\theta})}p_{\theta}(x_{n+1:T}|x_{1:n})p_{\theta}(y_{n+1:T}|y_{1:n},x_{1:T})\nonumber \\
 &  & \times\left\{ \prod_{k=2}^{n}\mathbb{I}[x_{1:k}\in\mathbf{s}_{k}]\right\} \nonumber \\
 & \propto & w_{n}^{\theta}(x_{1:n})p_{\theta}(x_{n+1:T}|x_{1:n})p_{\theta}(y_{n+1:T}|y_{1:n},x_{1:T})\nonumber \\
 & \propto & v_{n}^{\theta}\left(x_{1:n}\left|x_{n+1:T}\right.\right),\label{eq:backward_sampling_prop_conditional}\end{eqnarray}

\noindent where for the third proportionality we have used \eqref{eq:updateweight1_cond}-\eqref{eq:updateweight2_cond}
and an expansion of $W_{n}^{\theta}(x_{1:n})$ which is the direct
analogue of \eqref{eq:weightexpansion} but for final time index $n$. 

\noindent To establish the assertion of the proposition we use an
inductive argument over the iterations of the backward sampling algorithm
(indexed by $n=T,T-1,...,1$). The inductive hypothesis is that for
some index $n$ satisfying $1<m<n<T$ of the backward sampling procedure,
$\left(X_{1:T}',\mathbf{S}_{1},\mathbf{S}_{2},...,\mathbf{S}_{n}\right)$
obtained immediately after sampling from the backwards weights is
distributed according to the marginal distribution $\sum_{\mathbf{s}_{n+1}}...\sum_{\mathbf{s}_{T}}\pi_{\theta}^{N}(x_{1:T},\mathbf{s}_{1},\mathbf{s}_{2},...,\mathbf{s}_{T})$.
This implies $\left(X_{n:T}',\mathbf{s}_{1},\mathbf{s}_{2},...,\mathbf{s}_{n-1}\right)$
is distributed according to $\sum_{x_{1:n-1}}\sum_{\mathbf{s}_{n}}...\sum_{\mathbf{s}_{T}}\pi_{\theta}^{N}(x_{1:T},\mathbf{s}_{1},\mathbf{s}_{2},...,\mathbf{s}_{T})$.
Then at time step $n-1$, due to Eq. (\ref{eq:backward_sampling_prop_conditional}),
$\left(X_{1:T}',\mathbf{s}_{1},\mathbf{s}_{2},...,\mathbf{s}_{n-1}\right)$
obtained after sampling from the backward weights is distributed according
to $\sum_{\mathbf{s}_{n}}...\sum_{\mathbf{s}_{T}}\pi_{\theta}^{N}(x_{1:T},\mathbf{s}_{1},\mathbf{s}_{2},...,\mathbf{s}_{T})$
and thus $X_{1:T}'$ is distributed according to $\sum_{\mathbf{s}_{1}}...\sum_{\mathbf{s}_{T}}\pi_{\theta}^{N}(x_{1:T},\mathbf{s}_{1},\mathbf{s}_{2},...,\mathbf{s}_{T})=p_{\theta}(x_{1:T}|y_{1:T})$.
Next note that, due to Eq. (\ref{eq:backward_samp_prop_W}) the first
step of the backward sampling procedure draws from $\pi_{\theta}^{N}(x_{1:T}|\mathbf{s}_{1},\mathbf{s}_{2},...,\mathbf{s}_{T})$.
The proof is then complete under the assumption of the proposition.
$\blacksquare$

\noindent \textbf{Proof of Theorem} \ref{theorem:particlegibbs}.
For part $1$, it is easy to check that steps $1-4$ of the PG\
algorithm define a collapsed Gibbs sampler targeting Eq. (\ref{eq:artificialdistributiondetails}).
This follows from Proposition \ref{prop:backward_sampling} and the
fact that the conditional DPF update, given a value of $\theta$ and
$x_{1:T}$, is nothing but an algorithm sampling from \[
\left\{ \prod_{n=2}^{T}\mathbb{I}[x_{1:n}\in\mathbf{s}_{n}]\right\} \frac{\psi_{\theta}^{N}(\mathbf{s}_{1},\mathbf{s}_{2},...,\mathbf{s}_{T})}{\prod_{n=2}^{T}r_{\theta}^{N}(x_{1:n}\in\mathbf{s}_{n}|\mathbf{w}_{n-1}^{\theta})}.\]

For part $2$, we focus on establishing irreducibility and aperiodicity
of the transition probability of this algorithm. We denote by $\mathcal{L}_{G}$
the law of the Gibbs sampler to which assumption \ref{hyp:Gibbsconverges}
applies and $\mathcal{L}_{PG}^{N}$ the law of the PG sampler using
$N$ particles.

For any set $U$ write $2^{U}$ for the power set of $U$ and let
$\mathcal{B}(\Theta)$ denote a $\sigma$-algebra on $\Theta$. Let
$A\times B\times C\in\mathcal{B}(\Theta)\times2^{\mathcal{X}^{T}}\times\prod_{n=1}^{T}2^{\mathcal{P}(\mathcal{X}^{n})}$
be such that $\pi^{N}(\theta\in A,X_{1:T}\in B,\mathbf{S}_{1},...,\mathbf{S}_{T-1}\in C)>0$.
It follows that $\pi(\left(\theta,X_{1:T}\right)\in A\times B)>0$
and then from irreducibility of the corresponding Gibbs sampler (Assumption
\ref{hyp:Gibbsconverges}) there exists a finite $j$ such that $\mathcal{L}_{G}(\left(\theta(j),X_{1:T}(j)\right)\in A\times B)>0$.

From the definition of the conditional DPF update, it is straightforward
to check that, for any $\theta\in\Theta$, $N\geq2$, given any $x_{1:T}$
and for any time step, any particle which has positive weight immediately
before resampling has a positive probability of surviving that resampling
step. Thus, by an inductive argument in $n$, any point in the support
of $p_{\theta}(x_{1:T}|y_{1:T})$ has positive probability of being
assigned a positive weight at time $T$. It then follows from the
above arguments that $A\times B$ is marginally an accessible set
of the PG sampler for the same $j$: i.e. $\mathcal{L}_{PG}^{N}(\left(\theta(j),X_{1:T}(j)\right)\in A\times B)>0$.
Furthermore, as the conditional DPF update corresponds to drawing
from the conditional of $\pi^{N}$ given $\theta$ and $X_{1:T}$,
\[
\mathcal{L}_{PG}^{N}(\left(\theta(j+1),X_{1:T}(j+1),\mathbf{S}_{1}(j+1),...,\mathbf{S}_{T}(j+1)\right)\in A\times B\times C)>0\]
 and irreducibility follows. Furthermore, aperiodicity of the PG sampler
holds by contradiction: if the PG sampler were periodic, then the
Gibbs sampler would be too; this violates Assumption A\ref{hyp:Gibbsconverges}.$\blacksquare$

\bibliographystyle{Chicago}
\bibliography{switching}

\end{document}